\documentclass[preprint,aps]{revtex4}
\usepackage{graphicx}
\usepackage{epsfig}
\usepackage{CJK}
\usepackage{latexsym,amssymb,amsmath}
\usepackage{amsmath}
\usepackage{supertabular}
\usepackage{textcomp}
\usepackage{cases}

\begin{document}

\title{Quantum gravity corrections to the spontaneous excitation of an accelerated atom interacting with a quantum scalar field}
\author{Zhi Wang$^{}$$\footnote{E-mail address: zwangphys@163.com}$}

\affiliation{$^{}$School of Mathematics and Statistics, Guizhou University of Finance and Economics, Guiyang 550025, China}

\begin{abstract}

The Generalized Uncertainty Principle (GUP) extends the Heisenberg Uncertainty Principle (HUP) by suggesting a minimum observable scale that includes the effects of quantum gravity, which is supposed to potentially result in observable effects far below the Planck energy scale, providing us the opportunity to explore the theory of quantum gravity through physical processes at low energy scale. In present work, we study the corrections induced by the GUP to the spontaneous radiation properties of a two-level atom interacting with a real massless scalar quantum field based on the DDC formalism.
The GUP alters the correlation function of the scalar field, consequently affecting the radiative properties of atoms. We calculate the rate of change in the mean atomic energy for an atom undergoing inertial motion, uniform acceleration, and uniform circular motion.
We show that the GUP can modify the spontaneous emission rate of an excited-state atom in inertial motion; however, it does not alter the stability of the ground-state atom in vacuum. For an atom in uniformly accelerated and uniformly circular motions, the GUP can change both its spontaneous emission and excitation rates; moreover, the corrections caused by the GUP contains the terms proportional to $\beta a^2$ or $\beta a^3$, suggesting that the proper acceleration $a$ of an atom in non-inertial motions could significantly amplify the effect of the GUP on the spontaneous transition rates of the atom.

\par\vspace{\baselineskip}

\end{abstract}


\maketitle

\section{Introduction}

Over the past several decades, the reconciliation of quantum physics with general relativity stands as one of the paramount challenges in fundamental physics.
A favored strategy for this unification is quantization of gravity, but 
the theory of the quantized gravitational field would be non-renormalizable.
As of now, none of the proposals for quantum gravity have been experimentally validated, 
a complete description of quantum gravity remains elusive.
While several consistent features have emerged in all viable contenders for such a theory. One such feature is the presence of a minimal length scale at the Planck scale \cite{1,2}.

The string theory suggests that all of the different elementary particles stem from vibrating strings, 
where the string length represents the fundamental scale, and it is impossible to probe a scale smaller than its own length \cite{Amati,3,4,5}.
Gedanken experiments argue that the energy necessary to resolve spatial scales below the Planck length exceeds the energy sufficient to induce black hole formation within the probed region \cite{6,MM,7}.  
A minimum length is also a dynamic occurrence resulting from the constraint imposed by Planck length arising from quantum fluctuations of background gravitational field \cite{8,9}. 
In Doubly Special Relativity (DSR), the Lorentz symmetry is deformed, leading to an invariant energy scale. This deformation implies both a minimal length scale and a maximal momentum \cite{Camelia,10,11,12}.

However, the existence of a fundamental length scale contradicts the Heisenberg Uncertainty Principle (HUP), which suggests that the spatial resolution can be infinitely sharpened with sufficiently energetic probes. 
Incorporating minimal length into quantum mechanics requires extending the HUP to its generalized form, i.e., the Generalized Uncertainty Principle (GUP).
The introduction of this concept has garnered significant attention in recent decades, leading to a proliferation of literature exploring the modifications of GUP on a wide range of quantum phenomena \cite{14,Kempf,Ali,Pedram,15,Chung,16,17,Das,18,19,20,21,Maziashvili1,Maziashvili2,Vakili,22,23,24,25,26,27,28,29,30,31,32,33,34,35,Fadel,36,37,38,39,Artigas}. 
The potential experimental approaches, including microscopic \cite{40} and macroscopic harmonic oscillators \cite{41}, or using quantum optomechanics systems \cite{42,43,44,45}, have also been proposed.
In addition, the corrections to the Casimir effect based on several GUP proposals implying a minimal length were studied in Ref. \cite{46,47}.
Significant modifications to the Unruh effect have also been studied within the framework of GUPs \cite{48,49,50,51,52}. The GUP has also been extensively studied in the context of black holes or cosmology \cite{BH,BH0,BH1,BH2,BH3,BH4}.

Spontaneous emission, as a fundamental quantum process in light-matter interactions, has always been a subject of interest for many years. Previous studies have shown that this process can be ascribed to pure vacuum fluctuations \cite{53,54}, predominant radiation reaction \cite{55,56}, or combined effect of both \cite{57,58,59,60}. Milonni proposed that this interpretational ambiguity arises fundamentally from the freedom in ordering atomic and field operators within the Heisenberg picture \cite{58,59,60}. 
Notable advancements has been achieved by Dalibard, Dupont-Roc, and Cohen-Tannoudji (DDC), who put forward in Refs. \cite{61,62} that adopting a symmetric ordering between atomic and field operators is crucial to ensure the Hermiticity of contributions from both vacuum fluctuations and radiation reaction.  
According to the DDC prescription, for ground-state atoms, the impact of vacuum fluctuations and radiation reaction on the average rate of change of atomic energy perfectly counterbalance each other. This precise cancellation ensures that no transitions occur from the ground state, thus preserving the atom's stability. In contrast, for atoms initially in excited state, both contributions are equal in magnitude and identical in sign, jointly inducing a decrease in the mean atomic energy, leading to the so-called spontaneous emission.

Subsequently, Audretsch and M\"{u}ller \cite{63,631} expanded the DDC formalism to investigate how vacuum fluctuations and radiation reaction contribute to the spontaneous excitation rate of a two-level atom that is accelerating and interacting with a scalar field in free Minkowski spacetime. Their findings reveal that acceleration disrupts the balance between vacuum fluctuations and radiation reaction, suggesting that ground-state atoms can transition to excited states even in a vacuum. These insights not only corroborate the Unruh effect but also provide a compelling interpretation, as the spontaneous excitation of accelerated atoms can be seen as the fundamental physical mechanism behind the Unruh effect. In recent years, the DDC formalism has been widely applied to explore the spontaneous radiative properties of atoms in different backgrounds \cite{64,65,66,67,68,69,70,71,72,73,74,75,76,Barman}.

Considering the GUP-induced modifications of various quantum phenomena is universal, it is reasonable to expect that the GUP could also influence the spontaneous radiation properties of atoms.
By studying exquisitely sensitive systems such as atomic transitions, we can detect minuscule GUP-induced corrections. This offers a pathway to probe Planck-scale physics through tabletop experiments, shifting the search for quantum gravity from the abstract realm of cosmology into the laboratory.
In this paper, we aim to investigate the corrections caused by GUP to the spontaneous radiative processes of a two-level atom using the DDC formalism. Initially, the atom is prepared in an energy eigenstate, while the field is maintained in its vacuum state. We explore three distinct motion states of the atom, i.e., the uniform motion, uniform acceleration, and uniform circular motion. As a preliminary analysis, we consider the two-level atom to be weakly coupled to a bath of fluctuating real massless scalar quantum field.
The structure of the paper is as follows. In the subsequent section, the GUP proposal we adopted in present work and the Green's function in position space are brief reviewed. In Section III, the model of a two-level atom coupled to a scalar quantum field and the DDC formalism are introduced. In Section IV, we examine how the GUP alters the spontaneous emission of an atom in inertial motion. 
In Section V and VI, we extend this analysis to spontaneous excitation and de-excitation of an atom in non-inertial motions, with scenarios involving a uniformly accelerating atom and a uniformly circulating atom, respectively. 
The summary is given in last Section.

\section{The GUP proposal and Green's functions in position space}
The GUPs that implying a minimal length scale have been extensively studied in the past few decades, 
the pioneering works can be found in \cite{6,MM,14}.
Here we focus on the GUP model proposed by Kempf et al. \cite{14}, which has the form as
\begin{align}
	\Delta X\Delta P \ge \frac{\hbar }{2}(1 + \beta \Delta {P^2}),
\end{align}
where $ \beta=\beta_{0} / (M_{\mathrm{Pl}} c)^{2}=\beta_{0} l_{\mathrm{Pl}}^{2} / \hbar^{2}$ denotes the GUP parameter, with  $\beta_{0}$ being a dimensionless parameter assumed to be of order unity, and the Planck energy $M_{\mathrm{Pl}} c^{2} \simeq 10^{19} \mathrm{GeV}$, the Planck length $l_{\mathrm{Pl}} \simeq 10^{-35} \mathrm{~m}$. At energies significantly below the Planck energy, the $\beta$-dependent GUP correction becomes insignificant, leading to the recovery of HUP. 

It is clear that the uncertainty relation (1) corresponds to a minimum position uncertainty $\Delta {x_{\min }} \simeq {l _\mathrm{Pl}}\sqrt {{\beta _0}} $. For states exhibiting mirror symmetry, one can straightforward to derive the uncertainty relation (1) by use of the commutator:
\begin{align}
[X,P] = i\hbar (1 + \beta {P^2}).
\end{align}

The general form of the above expression for the three-dimensional scenario, retaining rotational isotropy, is provided by
\begin{align}
\left[X_{i}, P_{j}\right]=i\hbar \left(\delta_{i j}+\beta P^{2} \delta_{i j}+\beta^{\prime} P_{i} P_{j}\right) .
\end{align}
The position and momentum operators in the GUP framework still adhere to a Lie algebra structure. Therefore, the position commutator fixed by above equation and the Jacobi identity reads
\begin{align}
[{X_i},{X_j}] = i\hbar \frac{{2\beta  - \beta ' + (2\beta  + \beta ')\beta {P^2}}}{{1 + \beta {P^2}}}({P_j}{X_i} - {P_i}{X_j}).
\end{align}

As frequently done in the literature \cite{33,35}, we will consider the case $\beta ' = 2\beta$. By opting for this choice, the spatial geometric structure remains commutative up to $o(\beta ,\beta ')$, and then we get
$\left[X_{i}, P_{j}\right]=i\hbar \left(\delta_{i j}+\beta P^{2} \delta_{i j}+2 \beta P_{i} P_{j}\right)$,
the implementation of this algebra, up to the leading order in $\beta$, can be straightforwardly achieved using the usual position and momentum operators that satisfy $ \left[x_{i}, p_{j}\right]=i\hbar \delta_{i j}$, 
\begin{align}
X_{i}=x_{i}, \quad P_{i}=p_{i}\left(1+\beta {\mathbf{p}}^{2}\right) .
\end{align}
The Eq. (5) admits a simple physical interpretation that 
the momentum $ \mathbf{p} $ gains an increment of $ \beta \mathbf{p}^{2} \mathbf{p} $ because of quantum-gravitational fluctuations in the background field,
leading to the modified dispersion relation \cite{Maziashvili1,Maziashvili2}
\begin{align}
E^{2}=\mathbf{p}^{2}+m^{2}+2 \beta \mathbf{p}^{4} .
\end{align}

It is seen that the dispersion relation above, modified by the GUP, clearly violates Lorentz invariance. This GUP-induced modification causes photon propagation speeds to become energy-dependent, leading to the possibility of superluminal propagation. 
While superluminality appears unphysical, given that photons travel at the speed of light $c$ in vacuum according to Special Relativity, a reasonable assumption is that the principle of Special Relativity may not hold near or above the Planck scale $E_\mathrm{Pl}$ \cite{1}. Furthermore, it has been demonstrated that photons can travel subluminally or superluminally, depending on their trajectory through a gravitational field and the observer's position \cite{Lust}.
In the limit $\beta \to 0$, the standard dispersion relation with no quantum gravity correction is recovered.

The dispersion relation (6) leads to the GUP-modified propagator of
the scalar quantum field in position space as
\begin{align}
G\left(x, x^{\prime}\right)=\int_{}^{} \frac{\mathrm{d}^{4} p}{(2 \pi)^{4}} \frac{e^{-i\left[p_{0}\left(t-t^{\prime}\right)-\mathbf{p} \cdot\left(\mathbf{x}-\mathbf{x}^{\prime}\right)\right]}}{p_{0}^{2}-\mathbf{p}^{2}\left(1+2\beta \mathbf{p}^{2}\right)-m^{2}},
\end{align}
the $ p_{0} $ integral is performed by use of a contour integral, with the contour that corresponds to the relevant two-point correlation functions in the standard approach \cite{Birrell}. 
Then the positive frequency Wightman function modified by the GUP in the massless limit can be obtained as \cite{Davies}
\begin{align}
	D^{+}\left(x, x^{\prime}\right) =-\frac{1}{4 \pi ^{2}} \frac{1}{(\Delta t-i \varepsilon)^{2}-|\Delta \mathbf{x}|^{2}}\left(1-\frac{2 \beta}{(\Delta t-i \varepsilon)^{2}-|\Delta \mathbf{x}|^{2}}\right),
\end{align}
where the spacetime points are $x = (t, \mathbf{x})$ and an infinitesimally small positive
parameter $\varepsilon$ is introduced to characterize the singularities of the function.

\section{Atom-field interaction and The DDC formalism}

The system under consideration consists of a two-level atom weakly interacting with a real massless scalar quantum field. 
The total Hamiltonian of the system that governing its evolution in relation to the atom's proper time $\tau$ is given by
\begin{align}
H (\tau ) = {H_A}(\tau ) + {H_F}(\tau ) + {H_I}(\tau ),
\end{align}
where ${H_A}(\tau )=  {\omega _0}{R _3}(\tau )$ denotes Hamiltonian of the two-level atom, with $\omega _0$ being the energy level spacing of the atom, and $R_{3}=\frac{1}{2}(|+\rangle\langle +| - |- \rangle\langle-|)$. The units with $\hbar = c = 1$ is employed here and hereafter. ${H_F}(\tau )$ represents the Hamiltonian of the scalar field, which is expressed as
\begin{align}
H_{F}(\tau)=\int d^{3} k \omega_{\vec{k}} a_{\vec{k}}^{\dagger} a_{\vec{k}} \frac{d t}{d \tau}.
\end{align}
Through an analogy to electric dipole interaction, the atom and scalar field can be coupled as
${H_I}(\tau ) = \mu {R _2}(\tau ) \phi (x(\tau ))$ \cite{63}, with $\mu$ being a weak coupling constant, $R_{2}=\frac{i}{2} \left(R_{-}-R_{+}\right)$, where $ R_{+} = | +\rangle \langle -|$ and $R_{-} = | -\rangle \langle +| $ are the atomic raising and lowering
operators, respectively. $\phi (x)$ denotes the scalar field operator.
The coupling is effective solely along the atom's trajectory, $ x(\tau) $. 

The Heisenberg equations of motion for the dynamical variables associated with both the atom and the field can be derived with the above given Hamiltonian. 
To distinguish the contribution of vacuum field fluctuations and that of radiation reaction on the rate of change of atomic observables, we will analyze these two physical mechanisms separately. 
This is achieved by decomposing the field $\phi(x)$ into a ``free" component $\phi^{f}(x)$, which exists independently of atom-field interactions, and a ``source" component $\phi^{s}(x)$, arising from the interaction:  
$\phi(x) = \phi^{f}(x) + \phi^{s}(x)$.
Then the DDC formalism \cite{61,62} can be employed to isolate the influences of vacuum field fluctuations and radiation reaction on the evolution of atomic observables. Let's write out the Heisenberg equation for the atomic Hamiltonian
\begin{align}
\frac{d H_{A}(\tau)}{d \tau}=i \mu \omega_{0}\left[R_{2}(\tau), R_{3}(\tau)\right] \phi(x(\tau)),
\end{align}
by partitioning the field operator into the free component and the source component, and adopting a symmetric operator ordering between the variables related to the atom and the field, we derive the rate of change of atomic energy
\begin{align}
\frac{d H_{A}(\tau)}{d \tau}=\left(\frac{d H_{A}(\tau)}{d \tau}\right)_{v f}+\left(\frac{d H_{A}(\tau)}{d \tau}\right)_{r r},
\end{align}
with
\begin{align}
	\left(\frac{d H_{A}(\tau)}{d \tau}\right)_{v f} & =\frac{1}{2} i \mu \omega_{0}\left\{\phi^{f}(x(\tau)),\left[R_{2}(\tau), R_{3}(\tau)\right]\right\}, \\
	\left(\frac{d H_{A}(\tau)}{d \tau}\right)_{r r} & =\frac{1}{2} i \mu \omega_{0}\left\{\phi^{s}(x(\tau)),\left[R_{2}(\tau), R_{3}(\tau)\right]\right\}.
\end{align}

We assume that the atom is initially prepared in the state $\left|b\right\rangle$, while the field is in the vacuum state $\left|0\right\rangle$. By averaging the two equations above over the system's state $\left|0, b\right\rangle$, simplifying, we get the impacts of vacuum field fluctuations as well as atomic radiation reaction on the average rate of change of atomic energy
\begin{align}
	\left\langle\frac{\mathrm{d} H_{A}(\tau)}{\mathrm{d} \tau}\right\rangle_{vf} = 2 i \mu^{2} \int_{\tau_{0}}^{\tau} \mathrm{d} \tau^{\prime} C^{F }\left(x(\tau), x\left(\tau^{\prime}\right)\right) \frac{\mathrm{d}}{\mathrm{d} \tau} \chi_{b}^{A}\left(\tau, \tau^{\prime}\right), \\
	\left\langle\frac{\mathrm{d} H_{A}(\tau)}{\mathrm{d} \tau}\right\rangle_{rr} = 2 i \mu^{2} \int_{\tau_{0}}^{\tau} \mathrm{d} \tau^{\prime} \chi^{F }\left(x(\tau), x\left(\tau^{\prime}\right)\right) \frac{\mathrm{d}}{\mathrm{d} \tau} C_{b}^{A}\left(\tau, \tau^{\prime}\right),
\end{align}
where $C_{b}^{A}\left(\tau, \tau^{\prime}\right)$ and $\chi_{b}^{A}\left(\tau, \tau^{\prime}\right)$ are two statistical functions, called the symmetric correlation function and the linear susceptibility of the atom in the state $\left | b  \right \rangle$, respectively, whose expressions read
\begin{align}
	C_{b}^{A}\left(\tau, \tau^{\prime}\right) & = \frac{1}{2}\left\langle b\left|\left\{R_{2}^{f}(\tau), R_{2}^{f}\left(\tau^{\prime}\right)\right\}\right| b\right\rangle
	= \frac{1}{2} \sum_{d}\left|\left\langle b\left|R_{2}(0)\right| d\right\rangle\right|^{2}\left(e^{i \omega_{b d} \Delta \tau}+e^{-i \omega_{b d} \Delta \tau}\right), \\
	\chi_{b}^{A}\left(\tau, \tau^{\prime}\right) & = \frac{1}{2}\left\langle b\left|\left[R_{2}^{f}(\tau), R_{2}^{f}\left(\tau^{\prime}\right)\right]\right| b\right\rangle 
	= \frac{1}{2} \sum_{d}\left|\left\langle b\left|R_{2}(0)\right| d\right\rangle\right|^{2}\left(e^{i \omega_{b d} \Delta \tau}-e^{-i \omega_{b d} \Delta \tau}\right),
\end{align}
with $\omega_{b d}=\omega_b-\omega_d$ and $\Delta \tau= \tau- \tau'$, the sum spreads over a complete set of atomic stationary states.
$C^{F}\left(x(\tau), x\left(\tau^{\prime}\right)\right)$ and $\chi^{F}\left(x(\tau), x\left(\tau^{\prime}\right)\right)$ being the symmetric correlation function and the linear susceptibility of the field in vacuum state, respectively, are given by
\begin{equation}\begin{aligned}
&C^{F}\left(x(\tau), x\left(\tau^{\prime}\right)\right)=\frac{1}{2}\left\langle 0\left|\left\{\phi^{f}(x(\tau)), \phi^{f}\left(x\left(\tau^{\prime}\right)\right)\right\}\right| 0\right\rangle,\\
&\chi^{F}\left(x(\tau), x\left(\tau^{\prime}\right)\right)=\frac{1}{2}\left\langle 0\left|\left[\phi^{f}(x(\tau)), \phi^{f}\left(x\left(\tau^{\prime}\right)\right)\right]\right| 0\right\rangle.
\end{aligned}\end{equation}

Plugging the positive frequency Wightman function (8), the statistical functions of the field modified by the GUP can be written as
\begin{equation}\begin{aligned}
		C^{F}\left(x, x^{\prime}\right) 
		= & -\frac{1}{8 \pi^{2}}\left(\frac{1}{(\Delta t-i \varepsilon)^{2}-|\Delta \mathbf{x}|^{2}}+\frac{1}{(\Delta t+i \varepsilon)^{2}-|\Delta \mathbf{x}|^{2}}\right)+ \\
		& \frac{\beta}{4 \pi^{2}}\left(\frac{1}{\left((\Delta t-i \varepsilon)^{2}-|\Delta \mathbf{x}|^{2}\right)^{2}}+\frac{1}{\left((\Delta t+i \varepsilon)^{2}-|\Delta \mathbf{x}|^{2}\right)^{2}}\right), 
\end{aligned}\end{equation}	
\begin{equation}\begin{aligned}
		\chi^{F}\left(x, x^{\prime}\right)= & -\frac{1}{8 \pi^{2}}\left(\frac{1}{(\Delta t-i \varepsilon)^{2}-|\Delta \mathbf{x}|^{2}}-\frac{1}{(\Delta t+i \varepsilon)^{2}-|\Delta \mathbf{x}|^{2}}\right)+ \\
		& \frac{\beta}{4 \pi^{2}}\left(\frac{1}{\left((\Delta t-i \varepsilon)^{2}-|\Delta \mathbf{x}|^{2}\right)^{2}}-\frac{1}{\left((\Delta t+i \varepsilon)^{2}-|\Delta \mathbf{x}|^{2}\right)^{2}}\right),
\end{aligned}\end{equation}
where $\Delta t = t(\tau ) - t(\tau '),\Delta \mathbf{x} = \mathbf{x}(\tau ) - \mathbf{x}(\tau ')$.

\section{The uniformly moving atom}
In this section, we utilize the DDC formalism, previously described, to analyze the effect of GUP on the spontaneous emission of an inertial atom. This analysis will serve as a foundation for discussing the GUP-modified contributions of vacuum field fluctuations and radiation reactions in subsequent sections, which will address the more complex scenarios involving an accelerated atom. Considering an inertial atom moving in the $x$-direction with a constant velocity $\mathrm{v}$, whose trajectory reads
\begin{align}
t(\tau)=\gamma \tau, \quad \mathbf{x}(\tau)=\mathbf{x}_{0}+\mathbf{v} \gamma \tau,
\end{align}
where the Lorentz
factor $\gamma=(1-\mathbf{v}^2)^{-1/2}$. 
The statistical functions of the field can be easily obtained from the general forms Eqs. (20)
and (21) as
\begin{equation}\begin{aligned}
	C^{F}\left(x, x^{\prime}\right) = -\frac{1}{8 \pi^{2}}\left(\frac{1}{(\Delta \tau-i \varepsilon)^{2}}+\frac{1}{(\Delta \tau +i \varepsilon)^{2}}\right)+ \frac{\beta}{4 \pi^{2}}\left(\frac{1}{(\Delta \tau-i \varepsilon)^{4}}+\frac{1}{(\Delta \tau+i \varepsilon)^{4}}\right),
\end{aligned}\end{equation}	
\begin{equation}\begin{aligned}
	\chi^{F}\left(x, x^{\prime}\right) = -\frac{1}{8 \pi^{2}}\left(\frac{1}{(\Delta \tau-i \varepsilon)^{2}}-\frac{1}{(\Delta \tau+i \varepsilon)^{2}}\right)+ \frac{\beta}{4 \pi^{2}}\left(\frac{1}{(\Delta \tau-i \varepsilon)^{4}}-\frac{1}{(\Delta \tau+i \varepsilon)^{4}}\right),
\end{aligned}\end{equation}
where $\Delta \tau = \tau - \tau'$. Then by using the statistical
functions, we can evaluate 
the contribution of vacuum field fluctuations with the subscript ‘$vf$’
and that of radiation reaction with the subscript ‘$rr$’ to the average rate of change of atomic energy $\left\langle {H_A}(\tau ) \right\rangle $,
\begin{equation}\begin{aligned}
	{\left\langle {\frac{{d{H_A}(\tau )}}{{d\tau }}} \right\rangle _{vf}} = \frac{{{\mu ^2}}}{{8{\pi ^2}}}\sum\limits_d {{\omega _{bd}}{{\left| {\langle b|R_2^f(0)|d\rangle } \right|}^2}} \int_{ - \infty }^{ + \infty } {d\Delta \tau \left( {\frac{1}{{{{(\Delta \tau - i\varepsilon )}^2}}} + \frac{1}{{{{(\Delta \tau + i\varepsilon )}^2}}}} \right){e^{i{\omega _{bd}}\Delta \tau }}} \\
	- \frac{{\beta{\mu ^2}}}{{4{\pi ^2}}}\sum\limits_d {{\omega _{bd}}{{\left| {\langle b|R_2^f(0)|d\rangle } \right|}^2}} \int_{ - \infty }^{ + \infty } {d\Delta \tau \left( {\frac{1}{{{{(\Delta \tau - i\varepsilon )}^4}}} + \frac{1}{{{{(\Delta \tau + i\varepsilon )}^4}}}} \right){e^{i{\omega _{bd}}\Delta \tau }}},
\end{aligned}\end{equation}
\begin{equation}\begin{aligned}
	{\left\langle {\frac{{d{H_A}(\tau )}}{{d\tau }}} \right\rangle _{rr}} = \frac{{{\mu ^2}}}{{8{\pi ^2}}}\sum\limits_d {{\omega _{bd}}{{\left| {\langle b|R_2^f(0)|d\rangle } \right|}^2}} \int_{ - \infty }^{ + \infty } {d\Delta \tau \left( {\frac{1}{{{{(\Delta \tau - i\varepsilon )}^2}}} - \frac{1}{{{{(\Delta \tau + i\varepsilon )}^2}}}} \right){e^{i{\omega _{bd}}\Delta \tau }}} \\
	- \frac{{\beta{\mu ^2}}}{{4{\pi ^2}}}\sum\limits_d {{\omega _{bd}}{{\left| {\langle b|R_2^f(0)|d\rangle } \right|}^2}} \int_{ - \infty }^{ + \infty } {d\Delta \tau \left( {\frac{1}{{{{(\Delta \tau - i\varepsilon )}^4}}} - \frac{1}{{{{(\Delta \tau + i\varepsilon )}^4}}}} \right){e^{i{\omega _{bd}}\Delta \tau }}}. 
\end{aligned}\end{equation} 
Following some calculations, we determine the contribution of vacuum field fluctuations to the rate of change of $\left\langle {H_A}(\tau ) \right\rangle $ as
\begin{equation}\begin{aligned}
		\left\langle\frac{d H_{A}(\tau)}{d \tau}\right\rangle_{v f}= &
		-\frac{\mu^{2}}{4 \pi}\left[\sum_{\omega_{b}>\omega_{d}} 
		\omega_{b d}^{2}|\langle b|R_2^f(0)| d\rangle|^{2}
		- \sum_{\omega_{b}<\omega_{d}} 
		\omega_{b d}^{2}|\langle b|R_2^f(0)| d\rangle|^{2} \right ] \\
		& - \frac{\beta \mu^{2}}{12 \pi} \left[\sum_{\omega_{b}>\omega_{d}} 
		\omega_{b d}^{4}|\langle b|R_2^f(0)| d\rangle|^{2}
		- \sum_{\omega_{b}<\omega_{d}} 
		\omega_{b d}^{4}|\langle b|R_2^f(0)| d\rangle|^{2} \right ],
\end{aligned}\end{equation}
and that of atomic radiation reaction
\begin{equation}\begin{aligned}
		\left\langle\frac{d H_{A}(\tau)}{d \tau}\right\rangle_{r r}= &
		-\frac{\mu^{2}}{4 \pi}\left[\sum_{\omega_{b}>\omega_{d}} 
		\omega_{b d}^{2}|\langle b|R_2^f(0)| d\rangle|^{2}
		+ \sum_{\omega_{b}<\omega_{d}} 
		\omega_{b d}^{2}|\langle b|R_2^f(0)| d\rangle|^{2} \right ] \\
		& - \frac{\beta \mu^{2}}{12 \pi} \left[\sum_{\omega_{b}>\omega_{d}} 
		\omega_{b d}^{4}|\langle b|R_2^f(0)| d\rangle|^{2}
		+ \sum_{\omega_{b}<\omega_{d}} 
		\omega_{b d}^{4}|\langle b|R_2^f(0)| d\rangle|^{2} \right ] .
\end{aligned}\end{equation}

It is shown from Eqs. (27) and (28) that 
the corrections induced by the GUP are represented by $\beta$-dependent terms in the above results. And the effect of GUP only change both the contributions of vacuum field fluctuations as well as radiation reaction to
the rate of change of atomic energy quantitatively but not qualitatively since the GUP parameter $\beta > 0$. As $\beta \to 0$, our results reduce to those of the Minkowski spacetime with no quantum gravity correction \cite{63}. In addition, for a given atom, the radiation rate is always enhanced as compared with the case without GUP. 

From the Eq. (27), we see that for an atom initially in the excited state $\left| b \right\rangle  = \left|  +  \right\rangle$, 
one has $\left| d \right\rangle  = \left|  -  \right\rangle$, 
thus only the terms with ${\omega _b} > {\omega _d}$ contributes, in which situation $\left\langle\frac{d H_{A}(\tau)}{d \tau}\right\rangle_{v f} < 0$, this means that vacuum fluctuations tend to de-excite the atom in the excited state. 
While for an atom initially in the ground state $\left| b \right\rangle  = \left|  -  \right\rangle$, only the terms with ${\omega _b} < {\omega _d}$ survives and then $\left\langle\frac{d H_{A}(\tau)}{d \tau}\right\rangle_{v f} > 0$,
which implies that vacuum field fluctuations tend to excite atoms in their ground state.
 Note that if only contributions of vacuum fluctuations are considered, both spontaneous excitation and de-excitation would occur equally, no matter whether or not the GUP is taken into account.

On the other hand, Eq. (28) indicates that 
the radiation reaction always causes the atom to lose energy since $\left\langle\frac{d H_{A}(\tau)}{d \tau}\right\rangle_{r r} < 0$ for both the ground or excited state atom,
independent of whether the GUP effect is considered or not, 
leading to a problem of the instability of atoms. 
By incorporating the contributions of vacuum field
fluctuations and radiation reaction, the total average rate of change of the atomic energy can be expressed as
\begin{equation}\begin{aligned}
		\left\langle\frac{d H_{A}(\tau)}{d \tau}\right\rangle_{tot}=&
		\left\langle\frac{d H_{A}(\tau)}{d \tau}\right\rangle_{v f} 
		+ \left\langle\frac{d H_{A}(\tau)}{d \tau}\right\rangle_{r r}\\
		=& -\frac{\mu^{2}}{2 \pi} \sum_{\omega_{b}>\omega_{d}} 
		\omega_{b d}^{2}|\langle b|R_2^f(0)| d\rangle|^{2}
		- \frac{\beta \mu^{2}}{6 \pi} \sum_{\omega_{b}>\omega_{d}} 
		\omega_{b d}^{4}|\langle b|R_2^f(0)| d\rangle|^{2}.
\end{aligned}\end{equation}

We observe that for an atom in the ground state (${\omega _b} < {\omega _d}$),
the contributions of vacuum field fluctuations and atomic radiation reaction exactly cancel, 
in spite of with or without GUP considered. 
Hence, the GUP simultaneously alters the influence of vacuum field fluctuations and radiation reaction, such that the delicate balance between the two contributions shown in Ref. \cite{63}, in which the GUP is absent, remains.
Thus the effect of GUP does not alter the stability of ground-state inertial atoms
in vacuum. While for the excited state atom  $\left| b \right\rangle  = \left|  +  \right\rangle$, the effect of GUP 
can change the spontaneous emission rate of the atom, 
specifically, the second term of Eq. (29) proportional to $\beta \omega_{b d}^{4}$ is the correction induced by the GUP.

\section{The uniformly accelerated atom}

Next we generalize the discussion in above section to the
case of a uniformly accelerating atom.
We will study the effect of GUP on the spontaneous excitation for the atom interacting with a massless scalar quantum field.
Assuming the atom is being accelerated along the $z$ direction with proper acceleration $a$. The trajectory of the atom can be described by
\begin{align}
t(\tau)=\frac{1}{a} \sinh a \tau, \quad z(\tau)=\frac{1}{a} \cosh a \tau, \quad x(\tau)=y(\tau)=0.
\end{align}

Evaluating the GUP-modified symmetric correlation function (20) and linear susceptibility (21) of the scalar field along the trajectory (30), we obtain
\begin{equation}\begin{aligned}
	C^{F}\left(x(\tau), x\left(\tau^{\prime}\right)\right) &=-\frac{a^{2}}{32 \pi^{2}}\left[\frac{1}{\sinh ^{2}\left(\frac{a}{2}\left(\tau-\tau^{\prime}\right)-i a \epsilon\right)}+\frac{1}{\sinh ^{2}\left(\frac{a}{2}\left(\tau-\tau^{\prime}\right)+i a \epsilon\right)}\right] \\
	&+\frac{\beta a^{4}}{64 \pi^{2}} \left[\frac{1}{\sinh ^{4}\left(\frac{a}{2}\left(\tau-\tau^{\prime}\right)-i a \epsilon\right)}+\frac{1}{\sinh ^{4}\left(\frac{a}{2}\left(\tau-\tau^{\prime}\right)+i a \epsilon\right)}\right], \\
\end{aligned}\end{equation}	
\begin{equation}\begin{aligned}	
	\chi^{F}\left(x(\tau), x\left(\tau^{\prime}\right)\right) &=-\frac{a^{2}}{32 \pi^{2}}\left[\frac{1}{\sinh ^{2}\left(\frac{a}{2}\left(\tau-\tau^{\prime}\right)-i a \epsilon\right)}-\frac{1}{\sinh ^{2}\left(\frac{a}{2}\left(\tau-\tau^{\prime}\right)+i a \epsilon\right)}\right] \\
	&+\frac{\beta a^{4}}{64 \pi^{2}} \left[\frac{1}{\sinh ^{4}\left(\frac{a}{2}\left(\tau-\tau^{\prime}\right)-i a \epsilon\right)}-\frac{1}{\sinh ^{4}\left(\frac{a}{2}\left(\tau-\tau^{\prime}\right)+i a \epsilon\right)}\right].
\end{aligned}\end{equation}
 
After some calculations, we arrive at
\begin{equation}\begin{aligned}
		C^{F}\left(x(\tau), x\left(\tau^{\prime}\right)\right) =& - \left( {\frac{1}{{8{\pi ^2}}} + \frac{{\beta{a^2}}}{{24{\pi ^2}}}} \right)\sum\limits_{k =  - \infty }^\infty  {\left[ {\frac{1}{{{{\left( {\Delta \tau  + \frac{{2\pi i}}{a}k - 2i\epsilon} \right)}^2}}} + \frac{1}{{{{\left( {\Delta \tau  + \frac{{2\pi i}}{a}k + 2i\epsilon} \right)}^2}}}} \right]} \\
		& + \frac{{\beta}}{{4{\pi ^2}}}\sum\limits_{k =  - \infty }^\infty  {\left[ {\frac{1}{{{{\left( {\Delta \tau  + \frac{{2\pi i}}{a}k - 2i\epsilon} \right)}^4}}}} \right.} \left. { + \frac{1}{{{{\left( {\Delta \tau  + \frac{{2\pi i}}{a}k + 2i\epsilon} \right)}^4}}}} \right], 
\end{aligned}\end{equation}	
\begin{equation}\begin{aligned}
		\chi^{F}\left(x(\tau), x\left(\tau^{\prime}\right)\right) =& - \left( {\frac{1}{{8{\pi ^2}}} + \frac{{\beta{a^2}}}{{24{\pi ^2}}}} \right)\sum\limits_{k =  - \infty }^\infty  {\left[ {\frac{1}{{{{\left( {\Delta \tau  + \frac{{2\pi i}}{a}k - 2i\epsilon} \right)}^2}}} - \frac{1}{{{{\left( {\Delta \tau  + \frac{{2\pi i}}{a}k + 2i\epsilon} \right)}^2}}}} \right]} \\
		& + \frac{{\beta}}{{4{\pi ^2}}}\sum\limits_{k =  - \infty }^\infty  {\left[ {\frac{1}{{{{\left( {\Delta \tau  + \frac{{2\pi i}}{a}k - 2i\epsilon} \right)}^4}}}} \right.} \left. { - \frac{1}{{{{\left( {\Delta \tau  + \frac{{2\pi i}}{a}k + 2i\epsilon} \right)}^4}}}} \right].
\end{aligned}\end{equation}

Substituting the symmetric correlation function of the field (33) and antisymmetric statistical function of the atom (18) into Eq. (15), we get the contribution of vacuum field fluctuations to the average rate of change of the atomic excitation energy as
\begin{equation}\begin{aligned}
	\left\langle\frac{d H_{A}(\tau)}{d \tau}\right\rangle_{v f}= &
	 	\frac{{{\mu ^2}}}{{8{\pi ^2}}}\left( {1 + \frac{{\beta{a^2}}}{3}} \right)\sum\limits_d {{\omega _{bd}}} |\langle b|R_2^f(0)|d\rangle {|^2}\\
	 	& \times \sum\limits_{k =  - \infty }^\infty  {\int_{ - \infty }^\infty  d \Delta \tau } \left[ {\frac{1}{{{{\left( {\Delta \tau  + \frac{{2\pi i}}{a}k - 2i\epsilon} \right)}^2}}} + \frac{1}{{{{\left( {\Delta \tau  + \frac{{2\pi i}}{a}k + 2i\epsilon} \right)}^2}}}} \right]{e^{i{\omega _{bd}}\Delta \tau }}\\
	 	&- \frac{{\beta{\mu ^2}}}{{4{\pi ^2}}}\sum\limits_d {{\omega _{bd}}} |\langle b|R_2^f(0)|d\rangle {|^2}\\
	 	& \times \sum\limits_{k =  - \infty }^\infty  {\int_{ - \infty }^\infty  d \Delta \tau {\left[ {\frac{1}{{{{\left( {\Delta \tau  + \frac{{2\pi i}}{a}k - 2i\epsilon} \right)}^4}}}} \right.} \left. { + \frac{1}{{{{\left( {\Delta \tau  + \frac{{2\pi i}}{a}k + 2i\epsilon} \right)}^4}}}} \right]{e^{i{\omega _{bd}}\Delta \tau }}} .
\end{aligned}\end{equation}

Similarly, inserting the linear susceptibility of the field (34) and symmetric statistical function of the atom (17) into Eq. (16), the contribution of radiation reaction to the average rate of change of the atomic excitation energy is given by
\begin{equation}\begin{aligned}
	\left\langle\frac{d H_{A}(\tau)}{d \tau}\right\rangle_{r r}= &
		\frac{{{\mu ^2}}}{{8{\pi ^2}}}\left( {1 + \frac{{\beta{a^2}}}{3}} \right)\sum\limits_d {{\omega _{bd}}} |\langle b|R_2^f(0)|d\rangle {|^2}\\
		& \times \sum\limits_{k =  - \infty }^\infty  {\int_{ - \infty }^\infty  d \Delta \tau } \left[ {\frac{1}{{{{\left( {\Delta \tau  + \frac{{2\pi i}}{a}k - 2i\epsilon} \right)}^2}}} - \frac{1}{{{{\left( {\Delta \tau  + \frac{{2\pi i}}{a}k + 2i\epsilon} \right)}^2}}}} \right]{e^{i{\omega _{bd}}\Delta \tau }}\\
		&- \frac{{\beta{\mu ^2}}}{{4{\pi ^2}}}\sum\limits_d {{\omega _{bd}}} |\langle b|R_2^f(0)|d\rangle {|^2}\\
		& \times \sum\limits_{k =  - \infty }^\infty  {\int_{ - \infty }^\infty  d \Delta \tau {\left[ {\frac{1}{{{{\left( {\Delta \tau  + \frac{{2\pi i}}{a}k - 2i\epsilon} \right)}^4}}}} \right.} \left. { - \frac{1}{{{{\left( {\Delta \tau  + \frac{{2\pi i}}{a}k + 2i\epsilon} \right)}^4}}}} \right]{e^{i{\omega _{bd}}\Delta \tau }}}.
\end{aligned}\end{equation}
We can evaluate the integrals by use of
the residue theorem (The detailed derivation of Eqs. (35) and (36) is shown in Appendix A), and then obtain the analytical expression for the average rate of
change of atomic energy induced by vacuum fluctuation as
\begin{equation}\begin{aligned}
		\left\langle\frac{d H_{A}(\tau)}{d \tau}\right\rangle_{v f}= &
			- \frac{{{\mu ^2}}}{{4\pi }}\left( {1 + \frac{{\beta{a^2}}}{3}} \right)\left[ {\sum\limits_{{\omega _b} > {\omega _d}} {\omega _{bd}^2} {{\left| {\left\langle {b\left| {R_2^f(0)} \right|d} \right\rangle } \right|}^2}\left( {1 + \frac{2}{{{e^{2\pi {\omega _{bd}}/a}} - 1}}} \right)} \right.\\
			& \left. { - \sum\limits_{{\omega _b} < {\omega _d}} {\omega _{bd}^2} {{\left| {\left\langle {b\left| {R_2^f(0)} \right|d} \right\rangle } \right|}^2}\left( {1 + \frac{2}{{{e^{2\pi \left| {{\omega _{bd}}} \right|/a}} - 1}}} \right)} \right]\\
			&- \frac{{\beta{\mu ^2}}}{{12\pi }}\left[ {\sum\limits_{{\omega _b} > {\omega _d}} {\omega _{bd}^4} {{\left| {\left\langle {b\left| {R_2^f(0)} \right|d} \right\rangle } \right|}^2}\left( {1 + \frac{2}{{{e^{2\pi {\omega _{bd}}/a}} - 1}}} \right)} \right.\\
			& \left. { - \sum\limits_{{\omega _b} < {\omega _d}} {\omega _{bd}^4} {{\left| {\left\langle {b\left| {R_2^f(0)} \right|d} \right\rangle } \right|}^2}\left( {1 + \frac{2}{{{e^{2\pi \left| {{\omega _{bd}}} \right|/a}} - 1}}} \right)} \right].
\end{aligned}\end{equation}

From this result, we see that similar to the case of inertial atom, the vacuum fluctuation contributes to the average rate of change of the atomic energy in both ground state and excited state, which tend to excite the atom in the ground state and de-excite the atom in the excited state with the same amplitude, even though the effect of GUP is taken into account.
By comparing this result with that of inertial atom, we find that the contribution of vacuum fluctuation to the average energy change rate of uniformly accelerated atom has the thermal radiation terms related to the atomic acceleration. The $\beta$-dependent terms are the modification
induced by the GUP, 
this corrections change the rate of change of atomic energy significantly,
however, the thermal character is still retained.
In addition, the extra nonthermal term proportional to $\beta a^2$ has also appeared.
When $\beta \to 0$, we recover the
result obtained in Ref. \cite{63} for a uniformly accelerated atom in the usual Minkowski spacetime without the GUP.

Similarly, we have for the contribution of radiation reaction
\begin{equation}\begin{aligned}
		\left\langle\frac{d H_{A}(\tau)}{d \tau}\right\rangle_{r r}= &
			=  - \frac{{{\mu ^2}}}{{4\pi }}\left( {1 + \frac{{\beta{a^2}}}{3}} \right)\left[ {\sum\limits_{{\omega _b} > {\omega _d}} {\omega _{bd}^2} {{\left| {\left\langle {b\left| {R_2^f(0)} \right|d} \right\rangle } \right|}^2}} \right.\left. { + \sum\limits_{{\omega _b} < {\omega _d}} {\omega _{bd}^2} {{\left| {\left\langle {b\left| {R_2^f(0)} \right|d} \right\rangle } \right|}^2}} \right]\\
			& - \frac{{\beta{\mu ^2}}}{{12\pi }}\left[ {\sum\limits_{{\omega _b} > {\omega _d}} {\omega _{bd}^4} {{\left| {\left\langle {b\left| {R_2^f(0)} \right|d} \right\rangle } \right|}^2}} \right.\left. { + \sum\limits_{{\omega _b} < {\omega _d}} {\omega _{bd}^4} {{\left| {\left\langle {b\left| {R_2^f(0)} \right|d} \right\rangle } \right|}^2}} \right].
\end{aligned}\end{equation}

Comparing the Eqs. (28) with (38), we observe that when the effect of GUP is taken into consideration, the contribution of radiation reaction for a uniformly accelerated atom is obviously different with the case of the inertial atom, in which the Eq. (38) contains the $a$-dependent terms.
This is in contrast to the case of that without GUP \cite{63}, 
where the contribution of radiation reaction is not changed by acceleration, same as the case of a inertial atom, even for a uniformly accelerated one.
It suggests that, due to the effect of GUP, a uniformly accelerated atom on the trajectory
(30) would be subject to a radiation reaction force relying on the acceleration $a$.

The combined contributions of vacuum field fluctuations and radiation reaction yield the total rate of change for the atomic excitation energy as
\begin{equation}\begin{aligned}
		\left\langle\frac{d H_{A}(\tau)}{d \tau}\right\rangle_{tot}=&
		\left\langle\frac{d H_{A}(\tau)}{d \tau}\right\rangle_{v f} 
		+ \left\langle\frac{d H_{A}(\tau)}{d \tau}\right\rangle_{r r}\\
		=& - \frac{{{\mu ^2}}}{{2\pi }}\left( {1 + \frac{{\beta{a^2}}}{3}} \right)\left[ {\sum\limits_{{\omega _b} > {\omega _d}} {\omega _{bd}^2} {{\left| {\left\langle {b\left| {R_2^f(0)} \right|d} \right\rangle } \right|}^2}\left( {1 + \frac{1}{{{e^{2\pi {\omega _{bd}}/a}} - 1}}} \right)} \right.\\
		& \left. { - \sum\limits_{{\omega _b} < {\omega _d}} {\omega _{bd}^2} {{\left| {\left\langle {b\left| {R_2^f(0)} \right|d} \right\rangle } \right|}^2}\frac{1}{{{e^{2\pi \left| {{\omega _{bd}}} \right|/a}} - 1}}} \right]\\
		& - \frac{{\beta{\mu ^2}}}{{6\pi }}\left[ {\sum\limits_{{\omega _b} > {\omega _d}} {\omega _{bd}^4} {{\left| {\left\langle {b\left| {R_2^f(0)} \right|d} \right\rangle } \right|}^2}\left( {1 + \frac{1}{{{e^{2\pi {\omega _{bd}}/a}} - 1}}} \right)} \right.\\
		& \left. { - \sum\limits_{{\omega _b} < {\omega _d}} {\omega _{bd}^4} {{\left| {\left\langle {b\left| {R_2^f(0)} \right|d} \right\rangle } \right|}^2}\frac{1}{{{e^{2\pi \left| {{\omega _{bd}}} \right|/a}} - 1}}} \right].
\end{aligned}\end{equation}

We note that the distinct feature with the presence of the GUP is that the total rate
of change of the atomic mean energy now depends on not only GUP parameter $\beta$ but also the proper acceleration $a$ of the atom.
The terms proportional to $\beta a^{2}$ denote the nonthermal correction caused by GUP, and also suggest that the acceleration $a$ can amplify the GUP effect.
We see that for the atom initially prepared in the excited state, the spontaneous emission is modified by GUP and acceleration, with the appearance of both the thermal and nonthermal corrections as compared to an inertial atom, and although the GUP significantly affect the transition rate of the atom, the thermal character is still maintained.
However, for the atom initially prepared in the ground state, the
delicate balance between the vacuum fluctuations and radiation reaction 
is broken due to the uniformly accelerated motion as opposed to the inertial case, 
in spite of both contributions of the vacuum field fluctuations and radiation reaction are modified by the GUP, resulting in the transition of ground-state atom to excited occurred even in the vacuum, which is known as spontaneous excitation. 

From the Eq. (39), we can straightforwardly obtain the GUP-modified spontaneous excitation and emission rates, corresponding two Einstein coefficients $A_{\uparrow}$ and $A_{\downarrow}$, for the two-level atom with the energy gap $\omega_0$ between the ground state $| -\rangle$ and excited state $| +\rangle$ as
\begin{align}
	{A_ \uparrow } &= \frac{{{\mu ^2}{\omega _0}}}{{8\pi }}\left( {1 + \frac{{\beta {a^2}}}{3} + \frac{\beta }{3}\omega _0^2} \right)\frac{1}{{{e^{2\pi {\omega _0}/a}} - 1}},\\
	{A_ \downarrow } &= \frac{{{\mu ^2}{\omega _0}}}{{8\pi }}\left( {1 + \frac{{\beta {a^2}}}{3} + \frac{\beta }{3}\omega _0^2} \right)\left( {1 + \frac{1}{{{e^{2\pi {\omega _0}/a}} - 1}}} \right),
\end{align}
the ratio of the excitation and emission rates can give a thermal state
with an effective temperature $T_{\mathrm{eff}}$ defined from the detailed-balance Boltzmann factor, then we have
\begin{align}
	T_{\mathrm{eff}}=\omega_{0}\left[-\ln \left(A_{\uparrow} / A_{\downarrow}\right)\right]^{-1}.
\end{align}
It is seen that the effect of GUP always alters the transition rates by making
it as a function of $\beta$, 
however, the Unruh temperature for the atom undergoes uniformly accelerated motion does not changed by the GUP.
This result is consistent with that reported in Ref. \cite{Davies}, which was obtained using the standard DeWitt-Unruh detector method. We also note that \cite{48} investigated the radiation measured by an accelerated detector coupled to a scalar field in the presence of a fundamental minimal length. By evaluating the integral of the response rate associated with the minimal-length modified Wightman–Green function via the saddle-point approximation, the authors found that the net flux of field quanta is negligible, and thus no Planckian spectrum emerges.
In a related study, Ref. \cite{49} based on the form of the GUP proposed by Ali, Das, and Vagenas \cite{Ali}, explores how the GUP modifies the dispersion relation, the speed of photons, and the Unruh effect. Starting from the GUP-corrected dispersion relation, they derived the corresponding Klein–Gordon equation. After solving it to obtain the positive-frequency outgoing solution, they further derived the power spectrum measured by a uniformly accelerating observer. Their results demonstrate that the power spectrum includes a correction proportional to the GUP parameter 
$\alpha$. When this parameter vanishes, the spectrum reverts to the standard Planckian form.
Our results diverge from those in Refs. \cite{48} and \cite{49}. The reason lies in the modified dispersion relation and the corresponding Green's function used in our current work, which are different from those adopted in \cite{48,49}.

To estimate the acceleration required to produce GUP correction that comparable to the standard transition rates, we need to revert to the International System of Units and consider $ \beta=\beta_{0} / (M_{\mathrm{Pl}} c)^{2}$. Here, $\beta_{0}$ is a dimensionless parameter, and current experiments can predict larger upper bounds on it, with values generally ranging from $10^{30}$ to $10^{90}$ (see, for example, \cite{BH} and the references therein). This directly leads to the range of the order of magnitude of the acceleration being between $10^{10}$ and $10^{40}$ $\mathrm{cm/s^2}$.
When the $\beta \to 0$, the spontaneous excitation and emission rates will reduce to the results without the quantum gravitational corrections \cite{63}. Compared with the results shown in Ref. \cite{63}, the GUP significantly enhances the atomic transition rates by introducing additional terms that depend on the parameter $\beta$.
By subtracting the transition rates that does not take into account the GUP effect from Eqs. (40) and (41), we obtain the component purely induced by GUP as
\begin{align}
	{A^{\mathrm{GUP}}_ \uparrow } &= \frac{1}{3} \beta\omega _0^2 \gamma_0 \left( {1+\frac{{{a^2}}}{\omega _0^2} } \right)\frac{1}{{{e^{2\pi {\omega _0}/a}} - 1}},\\
	{A^{\mathrm{GUP}}_ \downarrow } &= \frac{1}{3} \beta\omega _0^2 \gamma_0 \left( {1+\frac{{{a^2}}}{\omega _0^2} } \right)\left( {1 + \frac{1}{{{e^{2\pi {\omega _0}/a}} - 1}}} \right),
\end{align}
where $\gamma_0=\frac{{{\mu ^2}{\omega _0}}}{{8\pi }}$ is the spontaneous emission rate for a two-level atom at rest.
Fig. 1 shows the ratios between GUP-induced contribution to the spontaneous excitation/emission rates and $\gamma_0$, i.e., $\frac{A^{\mathrm{GUP}}_ \uparrow}{\gamma_{0}}$ and $\frac{A^{\mathrm{GUP}}_ \downarrow}{\gamma_{0}}$, as a function of atomic acceleration for different values of the GUP parameter \(\beta\). 
We see the corrections caused by GUP to both spontaneous excitation and emission rates display monotonically increasing behaviors with the growth of the atomic acceleration, in which $A^{\mathrm{GUP}}_ \downarrow$ is slightly larger than $A^{\mathrm{GUP}}_ \uparrow$ for a same acceleration. 
This suggests that the acceleration of atom can significantly amplify the effect of the GUP on its transition rates.
For a fixed acceleration, both the two transition rates purely induced by GUP enhance with the increase of parameters $\beta$, meaning that the spontaneous transition rates can serve as the sensitive observable to the GUP. When the acceleration is sufficiently large, the two rates $A^{\mathrm{GUP}}_ \uparrow$ and $A^{\mathrm{GUP}}_ \downarrow$ tend to be the same.
\begin{figure}[htb]
	\centering
	\includegraphics[width=1.0\linewidth,angle=0,clip=true]{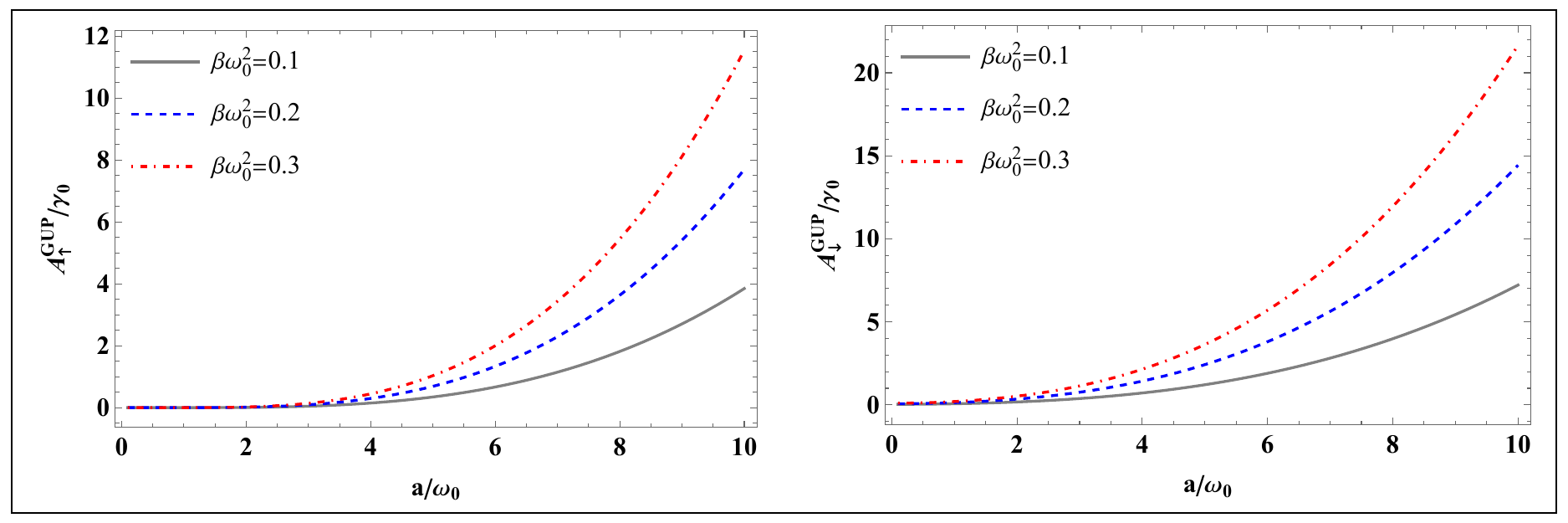}
	\caption{The behaviors of $\frac{A^{\mathrm{GUP}}_ \uparrow}{\gamma_{0}}$ (Left) and $\frac{A^{\mathrm{GUP}}_ \downarrow}{\gamma_{0}}$ (Right) for a uniformly accelerating atom with the increase of $\frac{a}{\omega _0}$. The solid, dashed, and dot-dashed lines refer to the cases for $\beta\omega _0^2=0.1$, $0.2$ and $0.3$, respectively.}
\end{figure} 
\section{The atom in circular motion} 
In this section, by use of the formalism presented in the preceding section, we turn to study the GUP effect on the spontaneous processes of a uniformly circulating atom whose trajectory is described by
\begin{align}
	x(\tau ) = \left( {\gamma \tau ,R\cos (\gamma \Omega \tau ),R\sin (\gamma \Omega \tau ), 0} \right),
\end{align}
where $R$ denotes orbital radius, $\Omega$ represents angular velocity in the preferred Lorentz frame, and $\gamma = (1 - v^2)^{-1/2}$ the Lorentz factor with $v = R\Omega$.
The proper acceleration $ a = R \Omega^{2} \gamma^{2} $.

Inserting the (45) into Eqs. (20) and (21), in the ultrarelativistic limit $\gamma  \gg 1$, we obtain the GUP-modified symmetric correlation function as well as the linear susceptibility of the field along the trajectory (45) as
\begin{equation}\begin{aligned}
	{C^F}(x,x') =& - \frac{1}{{8{\pi ^2}}}\left[ {\frac{1}{{{{(\Delta \tau  - i\varepsilon )}^2}\left( {1 + \frac{{{a^2}}}{{12}}\Delta {\tau ^2}} \right)}} + \frac{1}{{{{(\Delta \tau  + i\varepsilon )}^2}\left( {1 + \frac{{{a^2}}}{{12}}\Delta {\tau ^2}} \right)}}} \right] \\
	& + \frac{{\beta}}{{4{\pi ^2}}}\left[ {\frac{1}{{{{(\Delta \tau  - i\varepsilon )}^4}{{\left( {1 + \frac{{{a^2}}}{{12}}\Delta {\tau ^2}} \right)}^2}}} + \frac{1}{{{{(\Delta \tau  + i\varepsilon )}^4}{{\left( {1 + \frac{{{a^2}}}{{12}}\Delta {\tau ^2}} \right)}^2}}}} \right],
\end{aligned}\end{equation}

\begin{equation}\begin{aligned}
	{\chi ^F}(x,x') =  &- \frac{1}{{8{\pi ^2}}}\left[ {\frac{1}{{{{(\Delta \tau  - i\varepsilon )}^2}\left( {1 + \frac{{{a^2}}}{{12}}\Delta {\tau ^2}} \right)}} - \frac{1}{{{{(\Delta \tau  + i\varepsilon )}^2}\left( {1 + \frac{{{a^2}}}{{12}}\Delta {\tau ^2}} \right)}}} \right] \\
	&+ \frac{{\beta}}{{4{\pi ^2}}}\left[ {\frac{1}{{{{(\Delta \tau  - i\varepsilon )}^4}{{\left( {1 + \frac{{{a^2}}}{{12}}\Delta {\tau ^2}} \right)}^2}}} - \frac{1}{{{{(\Delta \tau  + i\varepsilon )}^4}{{\left( {1 + \frac{{{a^2}}}{{12}}\Delta {\tau ^2}} \right)}^2}}}} \right].
\end{aligned}\end{equation}

Through analogous derivation shown in the previous analysis, we derive the vacuum fluctuations contribution to the rate of change of the mean atomic energy as
\begin{equation}\begin{aligned}
	{\left\langle {\frac{{d{H_A}(\tau )}}{{d\tau }}} \right\rangle _{vf}} =  &- \frac{{{\mu ^2}}}{{4\pi }}\left[ {\sum\limits_{{\omega _b} > {\omega _d}} {{\omega _{bd}}{{\left| {\langle b|R_2^f(0)|d\rangle } \right|}^2}\left( {{\omega _{bd}} + \frac{a}{{2\sqrt 3 }}{e^{ - 2\sqrt 3 {\omega _{bd}}/a}}} \right)}  
 	    }\right.\\ & \left.{
	- \sum\limits_{{\omega _b} < {\omega _d}} {{\omega _{bd}}{{\left| {\langle b|R_2^f(0)|d\rangle } \right|}^2}\left( {{\omega _{bd}} - \frac{a}{{2\sqrt 3 }}{e^{2\sqrt 3 {\omega _{bd}}/a}}} \right)} } \right] \\
	& - \frac{{\beta{\mu ^2}}}{{4\pi }}\left\{ {\sum\limits_{{\omega _b} > {\omega _d}} {{\omega _{bd}}{{\left| {\langle b|R_2^f(0)|d\rangle } \right|}^2}\left[ {\frac{1}{3}{\omega _{bd}}\left( {{a^2} + {\omega _{bd}}^2} \right) + \left( {\frac{{5{a^3}}}{{24\sqrt 3 }} + \frac{{{a^2}{\omega _{bd}}}}{{12}}} \right){e^{ - 2\sqrt 3 {\omega _{bd}}/a}}} \right] 
  	    }}\right.\\ & \left.{{
	- \sum\limits_{{\omega _b} < {\omega _d}} {{\omega _{bd}}{{\left| {\langle b|R_2^f(0)|d\rangle } \right|}^2}\left[ {\frac{1}{3}{\omega _{bd}}\left( {{a^2} + {\omega _{bd}}^2} \right) - \left( {\frac{{5{a^3}}}{{24\sqrt 3 }} - \frac{{{a^2}{\omega _{bd}}}}{{12}}} \right){e^{2\sqrt 3 {\omega _{bd}}/a}}} \right]} } } \right\},
\end{aligned}\end{equation}
and the contribution caused by radiation reaction reads
\begin{equation}\begin{aligned}
	{\left\langle {\frac{{d{H_A}(\tau )}}{{d\tau }}} \right\rangle _{rr}} =  &- \frac{{{\mu ^2}}}{{4\pi }}\left[ {\sum\limits_{{\omega _b} > {\omega _d}} {{\omega _{bd}}^2{{\left| {\langle b|R_2^f(0)|d\rangle } \right|}^2}}  + \sum\limits_{{\omega _b} < {\omega _d}} {{\omega _{bd}}^2{{\left| {\langle b|R_2^f(0)|d\rangle } \right|}^2}} } \right]\\
	&- \frac{{\beta{\mu ^2}}}{{12\pi }}\left[ {\sum\limits_{{\omega _b} > {\omega _d}} {{\omega _{bd}}^2\left( {{a^2} + {\omega _{bd}}^2} \right){{\left| {\langle b|R_2^f(0)|d\rangle } \right|}^2}}  + \sum\limits_{{\omega _b} < {\omega _d}} {{\omega _{bd}}^2\left( {{a^2} + {\omega _{bd}}^2} \right){{\left| {\langle b|R_2^f(0)|d\rangle } \right|}^2}} } \right].
\end{aligned}\end{equation}

Obviously, both the two contributions are dependent on the GUP parameter $\beta$ and the acceleration $a$. 
Comparing (37) and (48) shows that the vacuum field fluctuations contribution to the average energy change rate diverges between uniformly circulating and uniformly accelerated atoms. Crucially, the characteristic thermal terms defined by the Planckian factor $\frac{1}{{{e^{2\pi {\omega _{bd}}/a}} - 1}}$, which depend on atomic acceleration, do not hold for the atom undergoing circular motion.
It is also found that the radiation reaction's contribution for a uniformly circulating atom (49) is exactly the same as that for a uniformly accelerated case (38). This result, which exhibits a clear acceleration dependence, stands in sharp contrast to the scenario where the GUP effect is not considered \cite{631}. In the limit $\beta \to 0$, the results
will reduce to the cases without the GUP corrections \cite{631}.
The total average rate of change of the atomic energy can be obtained as
\begin{equation}\begin{aligned}
	{\left\langle {\frac{{d{H_A}(\tau )}}{{d\tau }}} \right\rangle _{tot}} =  &- \frac{{{\mu ^2}}}{{4\pi }}\left[ {\sum\limits_{{\omega _b} > {\omega _d}} {{\omega _{bd}}{{\left| {\langle b|R_2^f(0)|d\rangle } \right|}^2}\left( {2{\omega _{bd}} + \frac{a}{{2\sqrt 3 }}{e^{ - 2\sqrt 3 {\omega _{bd}}/a}}} \right)}
	  }\right.\\ & \left.{
	+ \sum\limits_{{\omega _b} < {\omega _d}} {{\omega _{bd}}{{\left| {\langle b|R_2^f(0)|d\rangle } \right|}^2}\frac{a}{{2\sqrt 3 }}{e^{2\sqrt 3 {\omega _{bd}}/a}}} } \right] \\
	&- \frac{{\beta{\mu ^2}}}{{2\pi }}\left\{ {\sum\limits_{{\omega _b} > {\omega _d}} {{\omega _{bd}}{{\left| {\langle b|R_2^f(0)|d\rangle } \right|}^2}\left[ {\frac{{{\omega _{bd}}}}{3}\left( {{a^2} + {\omega _{bd}}^2} \right) + \left( {\frac{{5{a^3}}}{{48\sqrt 3 }} + \frac{{{a^2}{\omega _{bd}}}}{{24}}} \right){e^{ - 2\sqrt 3 {\omega _{bd}}/a}}} \right]}  
	    }\right.\\ & \left.{
	+ \sum\limits_{{\omega _b} < {\omega _d}} {{\omega _{bd}}{{\left| {\langle b|R_2^f(0)|d\rangle } \right|}^2}\left( {\frac{{5{a^3}}}{{48\sqrt 3 }} - \frac{{{a^2}{\omega _{bd}}}}{{24}}} \right){e^{2\sqrt 3 {\omega _{bd}}/a}}} } \right\}.
\end{aligned}\end{equation}

We see from (50) that for the atom in uniform circular motion,
both the spontaneous excitation and de-excitation are possible, and the effect of GUP can significantly modify the energy change rates of the atom.
To be specific, the uniformly circulating ground-state atom can be spontaneously excited and the transition rate for this process rely on both the acceleration $a$ and the GUP parameter $\beta$.
We also observe that the GUP-induced corrections contain the terms proportional to $\beta a^3$, indicating that the uniform circular motion of atom can also effectively amplify the GUP effect on the atomic transition rates.
In contrast to the case of linear acceleration, the terms associated with the Planckian factor in the total average rate of change of the atomic energy are substituted with terms that exhibit a non-Planckian exponential dependence. This means that the radiation perceived by an observer undergoing uniform circular motion is no longer thermal in nature.
The results return to the standard Minkowski spacetime case \cite{631} when $\beta$ approaches $0$, as expected.

To explicitly investigate the influence of the GUP on atomic transition rates, we extract the GUP-modified spontaneous excitation and emission rates from (50) as
\begin{align}
	{A_ \uparrow } &= \frac{{{\mu ^2}{\omega _0}}}{{8\pi }}\left[ {\frac{a}{{4\sqrt 3 {\omega _0}}}{e^{ - 2\sqrt 3 {\omega _0}/a}} + \beta \left( {\frac{{5{a^3}}}{{48\sqrt 3 {\omega _0}}} + \frac{{{a^2}}}{{24}}} \right){e^{ - 2\sqrt 3 {\omega _0}/a}}} \right],\\
	{A_ \downarrow } &= \frac{{{\mu ^2}{\omega _0}}}{{8\pi }}\left\{ {1 + \frac{a}{{4\sqrt 3 {\omega _0}}}{e^{ - 2\sqrt 3 {\omega _0}/a}} + \beta \left[ {\frac{1}{3}({a^2} + \omega _0^2) + \left( {\frac{{5{a^3}}}{{48\sqrt 3 {\omega _0}}} + \frac{{{a^2}}}{{24}}} \right){e^{ - 2\sqrt 3 {\omega _0}/a}}} \right]} \right\}.
\end{align}
\begin{figure}[htb]
	\centering
	\includegraphics[width=1.0\linewidth,angle=0,clip=true]{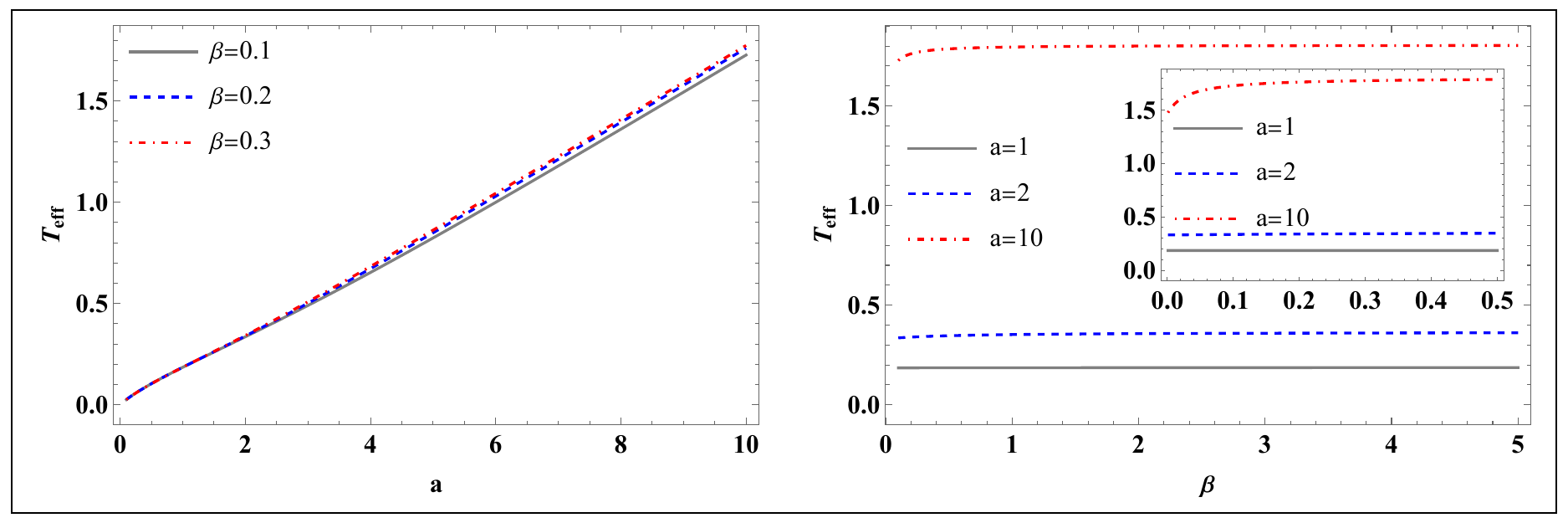}
	\caption{The GUP-modified effective temperature $T_{\mathrm{eff}}$ for the two-level atom undergoing the uniform circular motion as a function of the atomic acceleration (Left) and as a function of GUP parameter $\beta$ (Right).}
\end{figure} 
It is seen that the GUP effect consistently modifies the transition rates by introducing $\beta$-dependent terms. 
When assuming that $a$ is approximately equal to $\omega _0$, for the atom undergoing uniform circular motion, the order of magnitude of its acceleration required to induce the GUP correction comparable to the standard transition rate will be roughly the same as in the case of uniform accelerated linear motion,
while when $a$ is extremely large compared to $\omega _0$, the required acceleration for the uniformly circulating atom will be markedly reduced.
Even though the radiation detected by a circularly accelerated atom is non-thermal, as its transition rates lack the Planckian factor of thermal radiation, an effective temperature can still be defined with the Eq. (42). In Fig. 2, we plot the GUP-modified effective temperature for a two-level atom in uniform circular motion as a function of atomic acceleration as well as the GUP parameter $\beta$, where we set $\omega _0$ to 1.
From the left panel, we observe that the effective temperature $T_{\mathrm{eff}}$ increases with the acceleration. 
In contrast to the case of uniformly accelerated linear motion, the effective temperature here exhibits a dependence on the $\beta$. This dependence becomes particularly pronounced at high accelerations, where different values of $\beta$ lead to significant differences in the effective temperature.
When the acceleration is small, the temperature is less sensitive to different values of \( \beta \), whereas at higher accelerations, it becomes more sensitive to \( \beta \). That is, at larger accelerations, the effective temperature exhibits significant differences for different GUP parameter. 
This behavior is corroborated in the right panel: at low acceleration, the effective temperature remains largely insensitive to \( \beta \). In contrast, at high acceleration, it increases with \( \beta \) in the small-\( \beta \) regime and eventually saturates as \( \beta \) further increases.
It suggests that for an atom undergoing uniform circular motion, the GUP effect has a significant influence on the effective temperature perceived by the atom when the centripetal acceleration is sufficiently high.
We further isolate the purely GUP-induced contributions to the spontaneous excitation and emission rates as
\begin{align}
	{A^{\mathrm{GUP}}_ \uparrow } &= \frac{1}{24}\gamma_0 \beta\omega _0^2 { \left( {\frac{{5{a^3}}}{{2\sqrt 3 {\omega _0^3}}} + \frac{{{a^2}}}{{\omega _0^2}}} \right){e^{ - 2\sqrt 3 {\omega _0}/a}}},\\
	{A^{\mathrm{GUP}}_ \downarrow } &=\frac{1}{3} \gamma_0 \beta\omega _0^2  {\left[ {{1+\frac{{{a^2}}}{\omega _0^2} } + \left( {\frac{{5{a^3}}}{{16\sqrt 3 {\omega _0^3}}} + \frac{{{a^2}}}{{8\omega _0^2}}} \right){e^{ - 2\sqrt 3 {\omega _0}/a}}} \right]},
\end{align}
\begin{figure}[htb]
	\centering
	\includegraphics[width=1.0\linewidth,angle=0,clip=true]{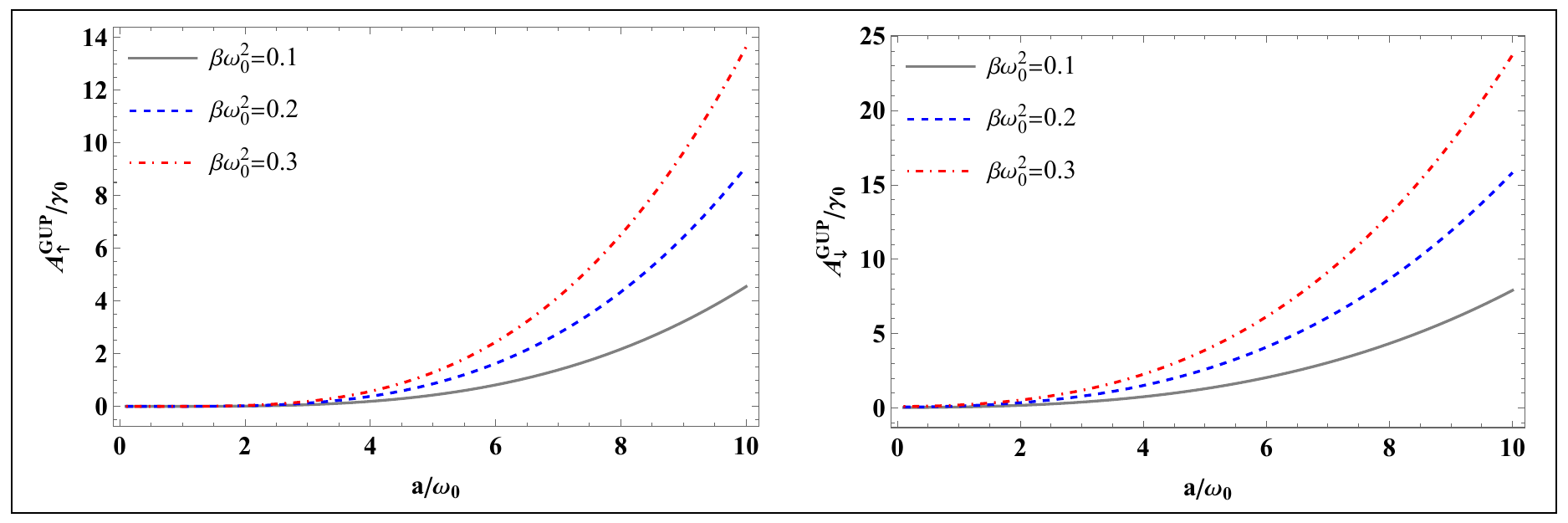}
	\caption{The behavior of $\frac{A^{\mathrm{GUP}}_ \uparrow}{\gamma_{0}}$ (Left) and $\frac{A^{\mathrm{GUP}}_ \downarrow}{\gamma_{0}}$ (Right) for a uniformly circulating atom with the growth of $\frac{a}{\omega _0}$. The solid, dashed, and dot-dashed lines refer to the cases for $\beta\omega _0^2=0.1$, $0.2$ and $0.3$, respectively.}
\end{figure} 

Shown in Fig. 3 is the variation of the ratio of GUP-induced contribution to the spontaneous excitation and emission rates to $\gamma_0$  as a function of atomic acceleration with different GUP parameter \( \beta \). 
Similar to the case of uniform accelerated linear motion, both $\frac{A^{\mathrm{GUP}}_ \uparrow}{\gamma_{0}}$ and $\frac{A^{\mathrm{GUP}}_ \downarrow}{\gamma_{0}}$ display monotonically increasing behaviors with the growth of the atomic acceleration, where $A^{\mathrm{GUP}}_ \downarrow$ is larger than $A^{\mathrm{GUP}}_ \uparrow$ for the same acceleration, and both the two rates purely caused by GUP increase with the parameters $\beta$ for a fixed acceleration.
As the acceleration grows, the difference in the GUP-induced corrections to the transition rates for different GUP parameter values becomes more pronounced.
Moreover, both \( A^{\mathrm{GUP}}_\uparrow \) and \( A^{\mathrm{GUP}}_\downarrow \) for the atom in uniform circular motion are larger than those for a uniformly accelerated atom for all values of the parameter \( \beta \).
This suggests that compared with uniformly accelerated linear motion, the uniform circular motion more effectively amplifies the effect of the GUP on both the spontaneous excitation and emission rates. That is, the transition rates of an atom undergoing uniform circular motion are more sensitive to the GUP.

\section{Summary}
We have studied the effect of the GUP on the spontaneous excitation and de-excitation for a two-level atom in interaction with a real massless scalar quantum field, and discussed the contributions of vacuum field fluctuations as well as radiation reaction to the rate of change of the mean energy of a two-level atom undergoing the inertial motion, uniform acceleration and uniform circular motions based on DDC formalism.

Our analysis reveals that for an inertial atom, the effect of GUP only change the contribution of vacuum field fluctuations as well as that of radiation reaction to the average rate of change of atomic energy quantitatively but not qualitatively.
For an atom initially in the ground state, the contributions of vacuum field fluctuations and radiation reaction exactly cancel, in spite of with or without GUP considered. Thus the GUP does not alter the stability of ground-state inertial atom in vacuum. For the excited state atom, the effect of GUP can enhance the spontaneous emission rate of the atom by adding a correction term proportional to $\beta \omega_{b d}^{4}$.

In the case of a uniformly accelerated atom, we show that the total rate of change of the mean atomic energy depends on both GUP parameter $\beta$ and the proper acceleration $a$ of the atom. 
The GUP not only quantitatively modified the thermal correction caused by acceleration, but also led to the emergence of non-thermal correction proportional to $\beta a^2$, suggesting that the acceleration could amplify the GUP effect.
For the atom initially prepared in the excited state, the spontaneous emission is modified by the GUP and atomic acceleration, arising of the $a$-dependent thermal and non-thermal terms as compared to the case of an inertial atom. For the atom initially prepared in the ground state, the transition from ground state to excited state known as spontaneous excitation is allowed to occur even in the vacuum. 
We also extract the GUP-modified spontaneous excitation and emission rates for the two-level atom and obtain the effective temperature from their ratio. It is found that while the GUP consistently alters the transition rates through the introduction of $\beta$-dependent terms, the effective temperature perceived by a uniformly accelerated atom remains unchanged. We further plot the ratios of the purely GUP-induced contributions to $\gamma_0$ as a function of atomic acceleration for different $\beta$. Both ratios increase monotonically with acceleration and are enhanced for growing $\beta$, indicating that the spontaneous transition rates can serve as a sensitive probe for the GUP at sufficiently high accelerations.

For the case of a uniformly circulating atom, in the ultrarelativistic limit $\gamma  \gg 1$, we observe the effect of GUP can significantly modify both the contributions of vacuum field fluctuations and radiation reaction to the average rates of change for the atomic energy. 
Our calculations reveal that the ground-state atom undergoing uniform circular acceleration exhibits spontaneous excitation with the transition rate determined by both the GUP parameter $\beta$ and atomic acceleration $a$.
Comparing the uniformly circulating and uniformly accelerated atomic motions, we observe distinct contributions from vacuum fluctuations. Specifically, the characteristic thermal terms defined by the Planckian factor, which depend on atomic acceleration, do not hold for circular motion. In contrast, the radiation reaction contribution is identical in both cases and exhibits a clear acceleration dependence. This behavior stands in sharp contrast to the scenario where the GUP effect is absent.
It is also seen that the corrections induced by GUP contain the terms proportional to $\beta a^3$, indicating that the uniform circular motion of atom can also significantly amplify the GUP effect.

We further obtain the GUP-modified spontaneous excitation and emission rates for the uniformly circulating atom, and plot the GUP-modified effective temperature as a function of atomic acceleration $a$ as well as the GUP parameter $\beta$. 
We see that in contrast to the case of uniformly accelerated linear motion, the effective temperature here exhibits a dependence on the $\beta$.
The effective temperature shows little sensitivity to the parameters at low acceleration, but becomes increasingly sensitive as the acceleration rises, particularly within the small parameter range. Furthermore, by comparing Fig.1 and Fig.3, we observe that for all different value of parameter \(\beta\), the GUP-modified rates \( A^{\mathrm{GUP}}_\uparrow \) and \( A^{\mathrm{GUP}}_\downarrow \) are consistently larger for uniform circular motion than for uniform linear acceleration, indicating a heightened sensitivity of the transition rates to the GUP in the circular case.
Thus it might provide a potentially way to allow us to probe the GUP experimentally if possible, and further deepen our understanding of quantum gravity and the nature of spacetime.

Finally, we note that the system studied here, a prepared two-level atom weakly coupled to a GUP-modified massless scalar quantum field, is to some extent a toy model. A more realistic system would involve a multilevel atom, such as a hydrogen atom, interacting with a quantized electromagnetic field.
Recently, we have investigated the GUP-induced corrections to the spontaneous excitation and de-excitation of a multilevel atom undergoing inertial and uniformly accelerated motions while coupled to a quantum electromagnetic field \cite{Wang1}. We observed corrections containing terms proportional to \(a^4\), suggesting that the GUP effect could be more effectively amplified by the atom's acceleration in this case than in the scalar field scenario.
Given that the extremely high acceleration required to observe the Unruh effect is challenging to achieve with linear motion, whereas large centripetal accelerations could be realized in certain settings, such as for ultrarelativistic electrons in storage rings. It is necessary to further study the GUP-modified spontaneous excitation and emission rates for a multilevel atom in uniform circular motion coupled to a quantum electromagnetic field.
Additionally, the results presented in this paper are based on the KMM form \cite{14} of the GUP. Other GUP proposals, for instance, those incorporating higher-order terms, different momentum-dependent functional forms, or models introducing both a minimum length and a maximum momentum, may lead to different forms of GUP corrections. Nevertheless, the core phenomena reported here, namely acceleration-dependent GUP corrections and the amplification of GUP effects by significant atomic acceleration, are expected to persist.

\section*{ACKNOWLEDGMENTS}
The work is supported by the Science and Technology Foundation of Guizhou Province (No. ZK[2022] General 029).

\section*{Appendix A: Derivation of Eqs. (35) and (36)}
\appendix          
\numberwithin{figure}{section}  
\setcounter{figure}{0}         
\renewcommand{\thefigure}{A\arabic{figure}} 

In this appendix, we will use the contour integration and residue theorem to calculate Eqs. (35) and (36), which contains four integrals, namely
\begin{equation}
\begin{array}{l}
	I_{1}=\int_{-\infty}^{\infty} \Phi_{1}(\Delta\tau) d \Delta \tau, \quad \Phi_{1}(\Delta\tau)=\frac{e^{i\omega_{b d}\Delta\tau}}{\left(\Delta\tau+\frac{2\pi i}{a}k-2i\varepsilon\right)^{2}}\\
	I_{2}=\int_{-\infty}^{\infty} \Phi_{2}(\Delta\tau) d \Delta \tau, \quad \Phi_{2}(\Delta\tau)=\frac{e^{i\omega_{b d}\Delta\tau}}{\left(\Delta\tau+\frac{2\pi i}{a}k+2i\varepsilon\right)^{2}} \\
	I_{3}=\int_{-\infty}^{\infty} \Phi_{3}(\Delta\tau) d \Delta \tau, \quad \Phi_{3}(\Delta\tau)=\frac{e^{i\omega_{b d}\Delta\tau}}{\left(\Delta\tau+\frac{2\pi i}{a}k-2i\varepsilon\right)^{4}} \\
	I_{4}=\int_{-\infty}^{\infty} \Phi_{4}(\Delta\tau) d \Delta \tau, \quad \Phi_{4}(\Delta\tau)=\frac{e^{i\omega_{b d}\Delta\tau}}{\left(\Delta\tau+\frac{2\pi i}{a}k+2i\varepsilon\right)^{4}}. \tag{A.1}
\end{array}
\end{equation}

We first compute the integral
\begin{align}
	I_1 = \int_{-\infty}^{\infty} \Phi_1(\Delta \tau) d\Delta \tau = \int_{-\infty}^{\infty} d\Delta \tau \frac{e^{i\omega_{b d} \Delta \tau}}{(\Delta \tau + \frac{2\pi i}{a} k - 2i\varepsilon)^2}. \tag{A.2}
\end{align}
In the above equation, the integrand \(\Phi_1(z)\) has a second-order pole at \(z_k = -\frac{2\pi i}{a} k + 2i\varepsilon\) in the complex plane, where \(k\) ranges from \(-\infty\) to \(\infty\). Considering the different possible values of \(\omega_{b d}\), we will compute the integral (A.2) for the cases \(\omega_{b d} > 0\) and \(\omega_{b d}< 0\) separately.

When \(\omega_{b d} > 0\), we can use the contour shown in Fig. A.1(a). The integrand \(\Phi_1(z)\) has second-order poles in the upper half-plane at \(z_k = -\frac{2\pi i}{a} k + 2i\varepsilon\), where $k$ ranges from \(-\infty\) to 0. Thus, according to the residue theorem, we have
\begin{align}
	\oint_C \Phi_1(z) dz = \int_{-\infty}^{\infty} \Phi_1(\Delta \tau) d\Delta \tau + \lim_{R \to \infty} \int_{C_R} \Phi_1(z) dz, \tag{A.3}
\end{align}
considering that $\lim_{R \to \infty} \int_{C_R} \Phi_1(z) dz = 0$. Therefore, we obtain
\begin{align}
	I_1 = \int_{-\infty}^{\infty} \Phi_1(\Delta \tau) d\Delta \tau = -2\pi \omega_{b d} e^{\frac{2\pi \omega_{b d}}{a} k}. \tag{A.4}
\end{align}

When \(\omega_{bd} < 0\), we can use the contour shown in Fig. A.1(b). The integrand \(\Phi_1(z)\) has second-order poles in the lower half-plane at \(z_k = -\frac{2\pi i}{a}k + 2i\varepsilon\), where \(k\) ranges from 1 to \(\infty\), then we directly arrive at
\begin{align}
	I_1 = \int_{-\infty}^{\infty} \Phi_1(\Delta\tau)d\Delta\tau = 2\pi \omega_{bd}e^{\frac{2\pi \omega_{bd}}{a}k}. \quad \tag{A.5}
\end{align}

Similarly, for the integral \(I_2\), we can also obtain via the residue theorem as
\begin{align}
	I_2 =
\begin{cases}
	-2\pi \omega_{bd}e^{-\frac{2\pi \omega_{bd}}{a}k}, & \text{if } \omega_{bd} > 0, k = -1, -2, -3, \ldots \\
	2\pi \omega_{bd}e^{\frac{2\pi \omega_{bd}}{a}k}, & \text{if } \omega_{bd} < 0, k = 0, 1, 2, \ldots
\end{cases}. \tag{A.6}
\end{align}
\begin{figure}[htb]
	\centering
	\includegraphics[width=1.0\linewidth,angle=0,clip=true]{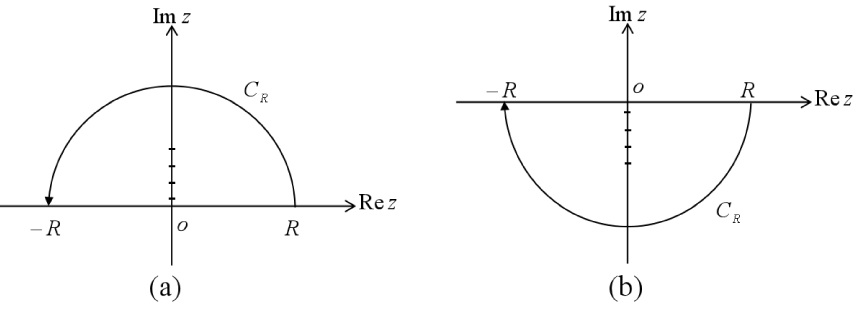}
	\caption{The Poles and integration paths.}
\end{figure}

Next, we compute the integral
\begin{align}
	I_3 = \int_{-\infty}^{\infty} \Phi_3(\Delta\tau) d\Delta\tau = \int_{-\infty}^{\infty} d\Delta\tau \frac{1}{(\Delta\tau + \frac{2\pi i}{a}k - 2i\varepsilon)^4} e^{i\omega_{bd}\Delta\tau}, \tag{A.7}
\end{align}
In the above equation, the integrand \(\Phi_3(z)\) has a fourth-order pole at \(z_k = -\frac{2\pi i}{a}k + 2i\varepsilon\) in the complex plane, where $k$ ranges from \(-\infty\) to \(\infty\). Considering the different possible values of \(\omega_{bd}\), we will compute the integral (A.7) for the cases \(\omega_{bd} > 0\) and \(\omega_{bd} < 0\) separately.

When \(\omega_{bd} > 0\), we can use the contour shown in Fig. A.1(a). The integrand \(\Phi_3(z)\) has fourth-order poles in the upper half-plane at \(z_k = -\frac{2\pi i}{a}k + 2i\varepsilon\), where $k$ ranges from \(-\infty\) to 0. Thus, by using the residue theorem, we have
\begin{align}
	\oint_C \Phi_3(z) dz = \int_{-\infty}^{\infty} \Phi_3(\Delta\tau) d\Delta\tau + \lim_{R \to \infty} \int_{C_R} \Phi_3(z) dz. \tag{A.8}
\end{align}
Considering that \(\lim_{R \to \infty} \int_{C_R} \Phi_3(z) dz = 0\). Therefore, we obtain
\begin{align}
	I_3 = \int_{-\infty}^{\infty} \Phi_3(\Delta\tau) d\Delta\tau = \frac{1}{3}\pi \omega_{bd}^3 e^{\frac{2\pi \omega_{bd}}{a}k}. \tag{A.9}
\end{align}

When \(\omega_{bd} < 0\), we can use the contour shown in Fig. A.1(b). The integrand \(\Phi_3(z)\) has fourth-order poles in the lower half-plane at \(z_k = -\frac{2\pi i}{a}k + 2i\varepsilon\), where \(k\) ranges from 1 to \(\infty\), Through the similar contour integration, we arrive at
\begin{align}
	I_3 = \int_{-\infty}^{\infty} \Phi_3(\Delta\tau)d\Delta\tau = -\frac{1}{3}\pi \omega_{bd}^3 e^{\frac{2\pi\omega_{bd}}{a}k}. \tag{A.10}
\end{align}

Similarly, for the integral \(I_4\), we can also obtain via the residue theorem:
\begin{align}
	I_4 = 
\begin{cases} 
	\frac{1}{3}\pi \omega_{bd}^3 e^{\frac{2\pi\omega_{bd}}{a}k}, & \text{if } \omega_{bd} > 0, k = -1, -2, -3, \ldots \\ 
	-\frac{1}{3}\pi \omega_{bd}^3 e^{\frac{2\pi\omega_{bd}}{a}k}, & \text{if } \omega_{bd} < 0, k = 0, 1, 2, \ldots 
\end{cases} . \tag{A.11}
\end{align}
Substituting the above integral results into Eqs. (35) and (36), and performing some algebraic calculations, we obtain Eqs. (37) and (38). The derivations for other cases, such as circular motion, follow similar derivation process and will not be elaborated further.


\begin{thebibliography}{99}

\bibitem{1} S. Hossenfelder, Minimal Length Scale Scenarios for Quantum Gravity, Living Rev. Relativity 16, 2 (2013). 
\bibitem{2} A. N. Tawfik and A. M. Diab, A review of the generalized uncertainty principle, Rep. Prog. Phys. 78, 126001 (2015). 

\bibitem{Amati} D. Amati, M. Ciafaloni, and G. Veneziano, Superstring Collisions at Planckian Energies, Phys. Lett. B 197 (1987) 81. 
\bibitem{3} D. Amati, M. Ciafaloni, G. Veneziano, Can spacetime be probed below the string size? Phys. Lett. B 216, 41 (1989). 
\bibitem{4} D. J. Gross, P. F. Mende, String theory beyond the Planck scale, Nucl. Phys. B 303, 407 (1988).
\bibitem{5} K. Konishi, G. Paffuti, P. Provero, Minimum physical length and the generalized uncertainty principle in string theory, Phys. Lett. B 234, 276 (1990). 
\bibitem{6} M. Maggiore, A generalized uncertainty principle in quantum gravity, Phys. Lett. B 304, 65 (1993). 
\bibitem{MM} M. Maggiore, The Algebraic structure of the generalized uncertainty principle, Phys. Lett. B 319 (1993) 83–86
\bibitem{7} M. Park, The generalized uncertainty principle in (A) dS space and the modification of Hawking temperature from the minimal length, Phys. Lett. B 659, 698 (2008). 
\bibitem{8} K. Nozari and T. Azizi, Some aspects of gravitational quantum mechanics, Gen. Rel. Grav. 38, 735 (2006). 
\bibitem{9} K. Nozari, Some aspects of Planck scale quantum optics, Phys. Lett. B 629, 41 (2005). 

\bibitem{Camelia} G. Amelino-Camelia, Testable scenario for relativity with minimum length, Phys. Lett. B 510, 255 (2001). 
\bibitem{10} J. Magueijo and L. Smolin, Lorentz invariance with an invariant energy scale, Phys. Rev. Lett. 88, 190403 (2002). 
\bibitem{11} J. Magueijo and L. Smolin, String theories with deformed energy-momentum relations, and a possible nontachyonic bosonic string, Phys. Rev. D 71, 026010 (2005). 
\bibitem{12} J. L. Cortes and J. Gamboa, Quantum uncertainty in doubly special relativity, Phys. Rev. D 71, 065015 (2005). 

\bibitem{14} A. Kempf, G. Mangano, R.B. Mann, Hilbert space representation of the minimal length uncertainty relation, Phys. Rev. D 52, 1108 (1995).
\bibitem{Kempf} A Kempf, On quantum field theory with nonzero minimal uncertainties in positions and momenta, J. Math. Phys. 38, 1347–1372 (1997).
\bibitem{Ali}A. F. Ali, S. Das, and E. C. Vagenas, Proposal for testing quantum gravity in the lab, Phys. Rev. D 84, 044013 (2011). 
\bibitem{Pedram} P. Pedram, A higher order GUP with minimal length uncertainty and maximal momentum, Phys. Lett. B 714, 317 (2012).
\bibitem{15}L. N. Chang, D. Minic, N. Okamura, T. Takeuchi, Exact solution of the harmonic oscillator in arbitrary dimensions with minimal length uncertainty relations, Phys. Rev. D 65.125027 (2002).
\bibitem{Chung} W. S. Chung, H. Hassanabadi, A new higher order GUP one dimensional quantum system, Eur. Phys. J. C 79, 213 (2019).
\bibitem{16}S. Benczik, L. N. Chang, D. Minic, T. Takeuchi, Hydrogen-atom spectrum under a minimal-length hypothesis, Phys. Rev. A 72, 012104 (2005). 
\bibitem{17}S. Haouat, Schrödinger equation and resonant scattering in the presence of a minimal length, Phys. Lett. B 729, 33 (2014).
\bibitem{Das}S. Das, R. B. Mann, Planck scale effects on some low energy quantum phenomena, Phys. Lett. B 704, 596 (2011). 
\bibitem{18}C. Villalpando, S. K. Modak, Minimal length effect on the broadening of free wave packets and its physical implications, Phys. Rev. D 100, 024054 (2019). 
\bibitem{19}P. Bosso, S. Das, R. B. Mann, Planck scale corrections to the harmonic oscillator, coherent, and squeezed states, Phys. Rev. D 96, 066008 (2017). 
\bibitem{20}J. Y. Bang, M. S. Berger, Quantum mechanics and the generalized uncertainty principle, Phys. Rev. D 74, 125012 (2006). 
\bibitem{21}P. Pedram, K. Nozari, S. H. Taheri, The effects of minimal length and maximal momentum on the transition rate of ultra cold neutrons in gravitational field, JHEP 03, 093 (2011). 

\bibitem{Maziashvili1} D. Mania, M. Maziashvili, Corrections to the black body radiation due to minimum-length deformed quantum mechanics, Phys. Lett. B 705, 521 (2011).
\bibitem{Maziashvili2} M. S. Berger and M. Maziashvili, Free particle wave function in light of the minimum-length deformed quantum mechanics and some of its phenomenological implications, Phys. Rev. D 84, 044043 (2011).

\bibitem{Vakili} B. Vakili and M. A. Gorji, Thermostatistics with minimal length uncertainty
relation, J. Stat. Mech. P10013 (2012).

\bibitem{22}A. Awad, A. F. Ali, Minimal length, Friedmann equations and maximum density, JHEP 06, 093 (2014).
\bibitem{23}X. Guo, P. Wang, H. Yang, The classical limit of minimal length uncertainty relation revisit with the Hamilton-Jacobi method, JCAP 05, 062 (2016).

\bibitem{24}L. B. Castro, A. E. Obispo, Generalized relativistic harmonic oscillator in minimal length quantum mechanics, J. Phys. A Math. Theor. 50, 285202 (2017).
\bibitem{25}M. I. Samar, V. M. Tkachuk, Exactly solvable problems in the momentum space with a minimum uncertainty in position uncertainty in position, J. Math. Phys. 57, 042102 (2016).
\bibitem{26}I. Prasetyo, I. H. Belfaqih, A. B. Wahidin, A. Suroso, A. Sulaksono, Minimal length, nuclear matter, and neutron stars, Eur. Phys. J. C 82, 884 (2022).
\bibitem{27}P. Bosso, G. G. Luciano, Generalized uncertainty principle from the harmonic oscillator to a QFT toy mode, Eur. Phys. J. C 81, 982 (2021).
\bibitem{28}T. L. Antonacci Oakes, R. O. Francisco, J. C. Fabris, J. A. Nogueira, Ground state of the hydrogen atom via Dirac equation in a minimal-length scenario, Eur. Phys. J. C 73, 2495 (2013).
\bibitem{29}F. J. Twagirayezu, Generalized uncertainty principle corrections on atomic excitation, Annals of Physics 422, 168294 (2020).
\bibitem{30}R. C. S. Bernardo, J. P. H. Esguerra, Energy levels of one-dimensional systems satisfying the minimal length uncertainty relation, Annals of Physics 373, 521 (2016).
\bibitem{31}F. Scardigli, R. Casadio, Generalized uncertainty principle, extra dimensions and holography, Class. Quantum Gravity 20, 3915 (2003). 
\bibitem{32} A. Camacho, Time Evolution of a Quantum Particle and a Generalized Uncertainty Principle, Rel. Grav.Cosmol. 1, 89 (2004). 
\bibitem{33}S. Das, E.C. Vagenas, Universality of quantum gravity corrections, Phys. Rev. Lett. 101, 221301 (2008). 
\bibitem{34} K. Nozari, P. Pedram, Minimal length and bouncing-particle spectrum, Europhys. Lett. 92, 50013 (2010). 
\bibitem{35} L. Petruzziello and F. Illuminati, Quantum gravitational decoherence from fluctuating minimal length and deformation parameter at the Planck scale, Nature Commun. 12, 4449 (2021). 
\bibitem{Fadel} M. Fadel, M. Maggiore, Revisiting the algebraic structure of the generalized uncertainty principle, Phys. Rev. D 105, 106017 (2022).
\bibitem{36}P. Bosso, L. Petruzziello, F. Wagner, F Illuminati, Spin operator, Bell nonlocality and Tsirelson bound in quantum-gravity induced minimal-length quantum mechanics, Communications Physics 6, 114 (2023). 
\bibitem{37}A. H. Gomes, On the algebraic approach to GUP in anisotropic space, Class. Quantum Grav. 40, 065005 (2023).
\bibitem{38}A. F. Ali, I. Elmashad, J. Mureika, Universality of minimal length, Phys. Lett. B 831, 137182 (2022).
\bibitem{39} R Casadio, W. Feng, I Kuntz, F. Scardigli, Minimum length (scale) in quantum field theory, generalized uncertainty principle and the non-renormalisability of gravity, Phys. Lett. B 838, 137722 (2023).
\bibitem{Artigas} D. Artigas, K. Martineau, J. Mielczarek, Squeezing of light from Planck-scale physics, Phys. Rev. D 109, 024028 (2024).

\bibitem{40} M. Bawaj et al., Probing deformed commutators with macroscopic harmonic oscillators, Nat. Commun. 6, 7503 (2015). 
\bibitem{41} F. Marin et al., Gravitational bar detectors set limits to Planck-scale physics on macroscopic variables, Nat. Phys. 9, 71 (2013). 
\bibitem{42} I. Pikovski, M. R. Vanner, M. Aspelmeyer, M. S. Kim, and Č. Brukner, Probing Planck-scale physics with quantum optics, Nat. Phys. 8, 393 (2012).
\bibitem{43} P. Bosso, S. Das, I. Pikovski, and M. Vanner, Amplified transduction of Planck-scale effects using quantum optics, Phys. Rev. A 96, 023849 (2017).
\bibitem{44} S. P. Kumar, M. B. Plenio, Quantum-optical tests of Planck-scale physics, Phys. Rev. A 97, 063855 (2018).
\bibitem{45} S. Sen, S. Bhattacharyya, S. Gangopadhyay, Probing the generalized uncertainty principle through quantum noises in optomechanical systems, Class. Quantum Grav. 39, 075020 (2022).

\bibitem{46}U. Harbach, S. Hossenfelder, The Casimir effect in the presence of a minimal length, Phys. Lett. B 632 (2006) 379-383.
\bibitem{47} A. M. Frassino, O. Panella, Casimir effect in minimal length theories based on a generalized uncertainty principle, Phys. Rev. D 85, 045030 (2012).
\bibitem{48} P. Nicolini, M. Rinaldi, A minimal length versus the Unruh effect, Phys. Lett. B 695, 303 (2011).
\bibitem{49} B. R. Majhi, E. C. Vagenas, Modified dispersion relation, photon's velocity, and Unruh effect, Phys. Lett. B 725, 477 (2013). 
\bibitem{50} V. Husain, J. Louko, Low energy Lorentz violation from modified dispersion at high energies, Phys. Rev. Lett. 116, 061301 (2016). 
\bibitem{51} Y. Gim, H. Um, W. Kim, Unruh effect of nonlocal field theories with a minimal length, Phys. Lett. B 784, 206 (2018).
\bibitem{52} F. Scardigli, M. Blasone, G Luciano, R Casadio, Modified Unruh effect from generalized uncertainty principle, Eur. Phys. J. C 78,728 (2018). 
\bibitem{BH} J. C. S. Neves, Upper bound on the GUP parameter using the black hole shadow, Eur. Phys. J. C 80, 343 (2020).
\bibitem{BH0} H. Chen, B. C. Lutfuolu, H. Hassanabadi and Z. Long, Thermodynamics of the Reissner Nordst-rom black hole with quintessence matter on the EGUP framework, Phys. Lett. B 827, 136994 (2022).
\bibitem{BH1} E. Sucu and İ. Sakallı, GUP-reinforced Hawking radiation in rotating linear dilaton black hole spacetime, Phys. Scr. 98, 105201 (2023).
\bibitem{BH2} E. Sucu, İ. Sakallı, Nonlinear electrodynamics effects on the geometry, thermodynamics, and quantum dynamics of (2 + 1)-dimensional black holes, Nucl. Phys. B 1015, 116894 (2025).
\bibitem{BH3} E. Sucu, İ. Sakallı, Quantum-corrected thermodynamics and plasma lensing of MOG black holes. Proc. R. Soc. A 481, 20250251 (2025).
\bibitem{BH4} C. Tekincay, G. Gecim and Y. Sucu, Zitterbewegung particles tunneling from Reissner-Nordström AdS black hole surrounded by quintessence, EPL 135, 31003 (2021).

\bibitem{53} T. A. Welton, Some observable effects of the quantum-mechanical fluctuations of the electromagnetic field, Phys. Rev. 74 1157 (1948).
\bibitem{54} G. Compagno, R. Passante and F. Persico, The role of the cloud of virtual photons in the shift of the ground state energy of a hydrogen atom, Phys. Lett. A 98 253 (1983).
\bibitem{55} J. R. Ackerhalt, P. L. Knight and J. H. Eberly, Radiation reaction and radiative frequency shifts, Phys. Rev. Lett. 30 456 (1973).
\bibitem{56} J. R. Ackerhalt and J. H. Eberly, Quantum electrodynamics and radiation reaction: nonrelativistic atomic frequency shifts and lifetimes, Phys. Rev. D 10 3350 (1974). 
\bibitem{57} I. R. Senitzky, Radiation-reaction and vacuum-field effects in Heisenberg-picture quantum electrodynamics, Phys. Rev. Lett. 31 955 (1973).
\bibitem{58} P. W. Milonni, J. R. Ackerhalt and W. A. Smith, Interpretation of radiative corrections in spontaneous emission, Phys. Rev. Lett. 31 958 (1973).
\bibitem{59} P. W. Milonni and W. A. Smith, Radiation reaction and vacuum fluctuations in spontaneous emission, Phys. Rev. A 11 814 (1975). 
\bibitem{60} P. W. Milonni, Semiclassical and quantum-electrodynamical approaches in nonrelativistic radiation theory, Phys. Rep. 25 1 (1976).
\bibitem{61} J. Dalibard, J. Dupont-Roc, and C. Cohen-Tannodji,Vacuum fluctuations and radiation reaction: identification of their respective contributions, J. Phys. (Paris) 43, 1617 (1982).
\bibitem{62} J. Dalibard, J. Dupont-Roc, and C. Cohen-Tannodji, Dynamics of a small system coupled to a reservoir: reservoir fluctuations and self-reaction, J. Phys. (Paris) 45, 637 (1984). 
\bibitem{63} J. Audretsch and R. Müller, Spontaneous excitation of an accelerated atom: The 
Contributions of vacuum fluctuations and radiation reaction, Phys. Rev. A 50 1755 (1994). 
\bibitem{631} J. Audretsch, R. Müller and M. Holzmann, Generalized Unruh effect and Lamb shift for atoms on arbitrary stationary trajectories, Class. Quantum Grav. 12, 2927 (1995).

\bibitem{64} H. Yu and S. Lu,Spontaneous excitation of an accelerated atom in a spacetime with a reflecting plane boundary, Phys. Rev. D 72 064022 (2005). 
\bibitem{65} Z. Zhu and H. Yu, Fulling–Davies–Unruh effect and spontaneous excitation of an accelerated atom interacting with a quantum scalar field, Phys. Lett. B 645 459 (2007).
\bibitem{66} Z. Zhu, H. Yu and S. Lu, Spontaneous excitation of an accelerated hydrogen atom coupled with electromagnetic vacuum fluctuations, Phys. Rev. D 73 107501 (2006). 
\bibitem{67} H. Yu and Z. Zhu, Spontaneous absorption of an accelerated hydrogen atom near a conducting plane in vacuum, Phys. Rev. D 74 044032 (2006). 
\bibitem{68} Y. Jin, J. Hu and H. Yu, Spontaneous excitation of a circularly accelerated atom coupled to electromagnetic vacuum fluctuations, Ann. Phys. 344 97 (2014).
\bibitem{69} W. Zhou and H. Yu, Interaction of Hawking radiation with static atoms outside a Schwarzschild black hole, JHEP 4, 024 (2007).  
\bibitem{70} Z. Zhu and H. Yu, Thermal nature of de Sitter spacetime and spontaneous excitation of atoms, JHEP 2, 033 (2008).
\bibitem{71} W. Zhou and H. Yu, Spontaneous excitation of a static multilevel atom coupled with electromagnetic vacuum fluctuations in Schwarzschild spacetime, Class. Quantum Grav. 29 085003 (2012).
\bibitem{72} G. Menezes, N. F. Svaiter, Radiative processes of uniformly accelerated entangled atoms, Phys. Rev. A 93, 052117 (2016).
\bibitem{73} H. Cai, Z. Wang, Z. Ren, Radiative properties of a static two-level atom in a cosmic dispiration spacetime, Class. Quantum Grav. 35, 155016 (2018).
\bibitem{74} S. Cheng, J. Hu, H. Yu, Spontaneous excitation of an accelerated atom coupled with quantum fluctuations of spacetime, Phys. Rev. D 100, 025010 (2019).
\bibitem{75} G. Menezes, Spontaneous excitation of an atom in a Kerr spacetime, Phys. Rev. D 95, 065015 (2017).
\bibitem{76} G. Picanço, N. F. Svaiter, C.A.D. Zarro, Radiative processes of entangled detectors in rotating frames, JHEP 08, 025 (2020).
\bibitem{Barman} S. Barman, B. R. Majhi, Radiative process of two entangled uniformly accelerated atoms in a thermal bath: a possible case of anti-Unruh event, JHEP 03, 245 (2021).

\bibitem{Lust} D. Lust, M. Petropoulos, Comment on superluminality in general relativity, Class. Quant. Grav. 29, 085013 (2012).
\bibitem{Birrell} N. D. Birrell and P. C. W. Davies, Quantum Fields in Curved Space, Cambridge University Press (1984).
\bibitem{Davies} P. C. W. Davies and P. Tee, Accelerated particle detectors with modified dispersion relations, Phys. Rev. D 108, 045009 (2023).
\bibitem{Wang1} Z. Wang and Z. Long, GUP corrections to the spontaneous excitation of an accelerated atom coupled with electromagnetic vacuum fluctuations, Phys. Lett. B 868, 139770 (2025).

\end{thebibliography}
\end{document}